\documentclass[paper,noblind,tablesonly,endfloat]{geophysics}

\usepackage{amssymb}
\usepackage{amscd}
\usepackage{amsfonts}
\usepackage{amsmath}
\usepackage{amssymb}
\usepackage{algorithm}
\usepackage{bm}
\usepackage{mathrsfs}
\usepackage{braket}
\usepackage{colortbl}
\usepackage{subfigure}
\usepackage{datetime}
\usepackage{fancyhdr}
\usepackage{float}
\usepackage{subfloat}
\usepackage{graphicx,caption}
\usepackage{makecell}
\usepackage{hyperref}
\usepackage[titletoc]{appendix}
\usepackage{nicefrac}
\usepackage[ntheorem]{empheq}
\usepackage{lscape}
\usepackage{epstopdf}
\usepackage{array}
\usepackage{algorithmic}
\usepackage{indentfirst}
\usepackage{lineno}
\usepackage[normalem]{ulem}

\newcommand{\comments}[1]{}

\begin{document}

\title{
Explicit coupling of acoustic and elastic wave propagation in finite difference simulations 
}

\ms{GEO-2019-0566} 

\address{
\footnotemark[1]{Formerly Division of Computer, Electrical and Mathematical Sciences and Engineering, King Abdullah University of Science and Technology, Thuwal 23955-6900, Saudi Arabia; presently Oden Institute for Computational Engineering and Sciences, The University of Texas at Austin, TX 78712, USA. E-mail: longfei.gao@kaust.edu.sa.}

\footnotemark[2]{Division of Computer, Electrical and Mathematical Sciences and Engineering, King Abdullah University of Science and Technology, Thuwal 23955-6900, Saudi Arabia. E-mail: david.keyes@kaust.edu.sa.}
}

\author{Longfei Gao\footnotemark[1] and David Keyes\footnotemark[2]}

\righthead{Coupled acoustic-elastic simulations}

\maketitle

\begin{abstract}
We present a mechanism to \textcolor{black}{explicitly} couple the \textcolor{black}{finite-difference} discretizations of 2D acoustic and isotropic elastic wave systems that are separated by straight interfaces.
Such coupled simulations allow the application of the elastic model to geological regions that are of special interest for seismic exploration studies (e.g., the areas surrounding salt bodies), while with the computationally more tractable acoustic model still being applied in the background regions. 
\textcolor{black}{
Specifically, the acoustic wave system is expressed in terms of velocity and pressure while the elastic wave system is expressed in terms of velocity and stress.
Both systems are posed in first-order forms and discretized on staggered grids.}
Special variants of the standard \textcolor{black}{finite-difference} operators, namely, operators that possess the summation-by-parts property, are used for the approximation of spatial derivatives. 
Penalty terms, which are also referred to as the simultaneous approximation terms, are designed to weakly impose the elastic-acoustic interface conditions in the \textcolor{black}{finite-difference} discretizations and couple the elastic and acoustic wave simulations together.
With the presented mechanism, we are able to perform the coupled elastic-acoustic wave simulations stably and accurately. 
Moreover, it is shown that the energy-conserving property in the continuous systems can be preserved in the discretization with carefully designed penalty terms.
\end{abstract}

\section{Introduction}
Waves propagating through earth media are routinely employed 
in seismic exploration. 
Many modern seismic imaging techniques require simulating seismic wave propagations repeatedly for numerous source terms and various instances of media parameters; 
see, for instance, \citet{symes2009seismic}, \citet{virieux2009overview}, and \citet{Schuster2011}. 

Since the true earth media are inaccessible to us, acoustic and elastic model assumptions are commonly used as surrogates in practice. 
\textcolor{black}{
More complex model assumptions exist, such as Biot's theory for porous media \textcolor{black}{\citep{biot1956theorya,biot1956theoryb}}, which can in theory provide more comprehensive depiction of the seismic waves, but is difficult to utilize effectively in current practice due to its expensive simulation cost and expanded number of model parameters.}
Between the acoustic and elastic media assumptions, 
the latter is considered more accurate for earth media since, 
in addition to the compressional wave supported by the former, 
it also supports shear and surface waves, which are commonly observed in seismic survey data.

However, the simulation cost associated with the elastic model is significantly higher than that of the acoustic model. 
Consider the following simple thought experiment involving a homogeneous medium with compressional wave-speed $c_p$ and shear wave-speed $c_s$, where $c_p \! > \! c_s$.
When simulating seismic waves, the spatial grid spacing is usually decided on a points per minimal wavelength basis. 
For a given frequency $f$, the minimal wavelength in elastic simulation (denoted as $\lambda_{\min}^E$) is determined by the smaller shear wave-speed as follows
\begin{equation}
\lambda_{\text{min}}^E = \frac{c_s}{f};
\end{equation}
the minimal wavelength in acoustic simulation (denoted as $\lambda_\text{min}^A$) is determined by the compressional wave-speed as follows
\begin{equation}
\lambda_\text{min}^A = \frac{c_p}{f},
\end{equation}
since the acoustic simulation concerns only the compressional wave.
For a prescribed number of grid points per minimal wavelength (denoted as $N_\text{ppw}$ hereafter), the resulting spatial grid spacings in elastic and acoustic simulations are 
\begin{equation}
\Delta^E_x = \frac{\lambda_\text{min}^E}{N_\text{ppw}} =  \frac{c_s}{f \cdot N_\text{ppw}} 
\quad\ \text{and} \quad\ 
\Delta^A_x = \frac{\lambda_\text{min}^{\textcolor{black}{A}}}{N_\text{ppw}} =  \frac{c_p}{f \cdot N_\text{ppw}}\ ,
\end{equation} 
respectively, which leads to the ratio $\nicefrac{\Delta^A_x}{\Delta^E_x} = \nicefrac{c_p}{c_s}$.
Furthermore, since seismic wave simulations usually employ explicit time stepping methods, 
the time step lengths are usually dictated by the Courant-Friedrichs-Lewy (CFL) stability condition. In short, we have
\begin{equation}
\Delta_t^E \propto \frac{\Delta_x^E}{c_p}
\quad\ \text{and} \quad\ 
\Delta_t^A \propto \frac{\Delta_x^A}{c_p}\ ,
\end{equation}
where $\Delta_t^E$ and $\Delta_t^A$ stand for the time step lengths in elastic and acoustic simulations, respectively. 
Assuming that the Courant numbers for both cases are the same, we then have the ratio $\nicefrac{\Delta^A_t}{\Delta^E_t} = \nicefrac{\Delta^A_x}{\Delta^E_x} = \nicefrac{c_p}{c_s}$. 

Taking the ratio $\nicefrac{c_p}{c_s}$ to be 2,\footnotemark[3] then, with three spatial dimensions and one temporal dimension, the overall space-time discretization points in the elastic simulation can be 16 times as many as that of the acoustic simulation. 
Furthermore, when posed as first-order systems in terms of velocity and pressure (acoustic case) or velocity and stress (elastic case), the spatial derivatives appearing in the isotropic elastic system are \textcolor{black}{three} times as many as that of the acoustic system; approximation of these spatial derivatives (e.g., via applying stencils) represents majority of the computations during seismic wave simulations.
Altogether, the ratio of computational costs associated with these two models reaches almost 50 in this simple thought experiment. 
\footnotetext[3]{The ratio $\nicefrac{c_p}{c_s} = 2$ is common for fluid saturated sediments; see, for instance, \cite{gregory1976fluid}. 
Moreover, according to \cite{hamilton1979v}, exceptionally high $\nicefrac{c_p}{c_s}$ ratios (above 13) can appear in sea floor sediments.
}

The above observation motivates this study on coupling elastic and acoustic models for seismic wave simulations so that application of the more expensive elastic model can be restricted to regions where it makes significant impacts.
One such region is the area surrounding a salt body, which is of particular interest in seismic exploration since salt bodies often form caps for oil and gas reservoirs.
The contrast in wave-speed values between a salt body and its surrounding sediments can be significant. 
For instance, the salt body can have compressional wave-speed values above 4500~m/s and shear wave-speed values above 2500~m/s with its surrounding sediments having compressional wave-speed values around 2000~m/s \textcolor{black}{\citep[see][]{thierry1987acoustics, leveille2011subsalt, jackson2017salt}}.
This high contrast in wave-speeds often leads to strong wave conversions at salt boundaries, which can be problematic for seismic imaging and interpretation under the acoustic assumption \textcolor{black}{\citep[see][]{ogilvie1996effects, lu2003identifying, jones2014seismic}}, among others. 

\textcolor{black}{
Specifically, when entering the salt body, the downward propagating P wave can split into transmitted P and S waves at the top of salt and continue their paths inside the salt body.
Upon reflection at the base of salt, both transmitted waves can return upward as P or S waves.
When exiting the salt body, the base-of-salt reflected upward propagating S wave can convert to P wave again at the top of salt.
In total, four possible wave paths exist for waves entering and exiting the salt body as P wave, denoted as PPPP, PPSP, PSPP, and PSSP following the nomenclature from \cite{ogilvie1996effects}\footnotemark[4].}
\footnotetext[4]{
\textcolor{black}{
The first and last letters indicate the types of downward and upward propagating waves outside the salt body;
the two letters in the middle indicate the types of downward and upward propagating waves inside the salt body.
In conventional seismic imaging practice, only P wave events are exploited, hence the four considered wave paths all start and end with the letter `P'. 
}}

\textcolor{black}{
In the presence of salt, all four wave paths mentioned above can carry
significant energy \textcolor{black}{(see Fig.~7 of \citealp{ogilvie1996effects} and Fig.~17 of \citealp{lu2003identifying})}.
If migrated with the P wave velocity, the three wave paths containing converted waves (i.e., PPSP, PSPP, and PSSP) can obscure or misinform the sub-salt structures. 
On the other hand, they have the potential to be used constructively to corroborate or complement the base-of-salt image formed by the PPPP events alone. 
Elastic modeling in the region surrounding the salt body is crucial to correctly capture these converted waves and to facilitate the appropriate subsequent processing steps, may it be identifying and removing, or constructively exploiting these converted wave information.}

A simple approach to couple the elastic and acoustic wave simulations is to use the elastic wave equation everywhere and simply set the shear wave-speed outside the interested region to zero.
There are \textcolor{black}{several} drawbacks with this approach. 
First, it leads to unnecessary computations outside the interested region. 
Second, it introduces abrupt jumps in the shear wave-speed, which often requires finer discretization to resolve \textcolor{black}{\citep[see][]{brown1984note, symes2009interface}.}
\textcolor{black}{
Third, certain solution components, such as the particle velocity tangential to the interface, are assumed to be continuous across the interface while the physical interface conditions do not pose such constraint \citep[see][]{singh2019coupled}, which is similar to the situation in fault modeling \citep[see][]{rojas2008modelling}.}
In this study, we present a mechanism to \textcolor{black}{explicitly} 
couple 2D acoustic and isotropic elastic wave simulations that are separated by straight interfaces.
Geometric configuration of the abstracted simulation domain 
can be found
in Figure \textcolor{black}{\ref{Domain_illustration}}.
Although we only address the 2D case in this study, the same methodology can be extended to the 3D case, since the presented interface treatment only requires modifications along the perpendicular directions of the interfaces.

Earlier works on explicitly coupling elastic and acoustic wave simulations can be found in, e.g., \citet{komatitsch2000wave}, \citet{chaljub2003solving}, \citet{kaser2008highly}, \citet{michler2009numerical}, \citet{wilcox2010high}, \citet{rodriguez2016non}, \citet{ye2016discontinuous}, \citet{appelo2018energy}, \citet{terrana2018spectral}, which employ the finite element discretization approach. 
On the other hand, within the \textcolor{black}{finite-difference} discretization approach, some earlier works on explicitly coupling elastic and acoustic wave simulations can be found in \citet{stephen1983comparison} and \citet{stephen1985finite}, where the elastic-acoustic interface conditions are used to set up linear systems to deduce the formulas for solution updates at the interfaces.
\textcolor{black}{
Some recent works can be found in \citet{singh2019coupled} and \citet{qu2020fluid}, which resort to the mimetic \textcolor{black}{finite-difference} approach. 
Interested readers may consult \citet{rojas2008modelling}, \citet{castillo2013mimetic}, \citet{de2014mimetic}, \citet{shragge2017solving} and the references therein for more information about the mimetic discretization technique.
}

Due to its simplicity and cost-effectiveness, the \textcolor{black}{finite-difference} discretization approach remains the popular choice in seismic exploration, with strong interest in furthering its development. 
In this study, we adopt the so-called SBP-SAT technique for spatial discretization, where SBP stands for summation-by-parts and refers to the property of the \textcolor{black}{finite-difference} operators while SAT stands for simultaneous approximation term, which is the penalty term introduced to address boundary or interface conditions.
These concepts date back to \cite{kreiss1974finite} and \cite{carpenter1994time}, respectively. 
\textcolor{black}{
Interested readers may consult the two review papers \citet{fernandez2014review} and \citet{svard2014review}, as well as the references therein for more details.
Other more recent related works include \citet{kozdon2014constraining}, \citet{lotto2015high}, \citet{o2017energy}, \citet{banks2018galerkin}, \citet{petersson2018high}, \citet{gao2019combining}.
}

In terms of the procedure, the respective wave systems on acoustic and elastic regions are first discretized using SBP \textcolor{black}{finite-difference} operators without concern for the interface conditions.
These semi-discretized wave systems are then coupled together through penalty terms, i.e., SATs, that impose the interface conditions weakly.
With carefully designed penalty terms, we show that the overall semi-discretization preserves the energy-conserving property in the continuous systems.
We believe that this particular discretization approach using SBP \textcolor{black}{finite-difference} operators and weak imposition of the interface conditions with the energy-conserving property  built in has not been presented for the 
coupled simulation of elastic-acoustic wave system 
yet and is worth sharing with the seismic community.

The rest of this paper is organized as follows. 
We first briefly describe the geometric configuration of the simulation domain and the acoustic and isotropic elastic wave systems under consideration.
We then present the semi-discretizations on both acoustic and elastic regions using SBP \textcolor{black}{finite-difference} operators, respectively, as well as the penalty terms that couple them together.
Finally, we demonstrate the efficacy of the proposed interface treatment with numerical examples.

\section{Problem Description} \label{Section_problem_description}
\begin{figure}[H]
\captionsetup{width=1\textwidth, font=small,labelfont=small}
\centering
\hspace{-0.8em}
\subfigure[\!\!)]{
\captionsetup{width=1\textwidth, font=small,labelfont=small}
\centering\includegraphics[width=0.45\textwidth, scale=0.125]{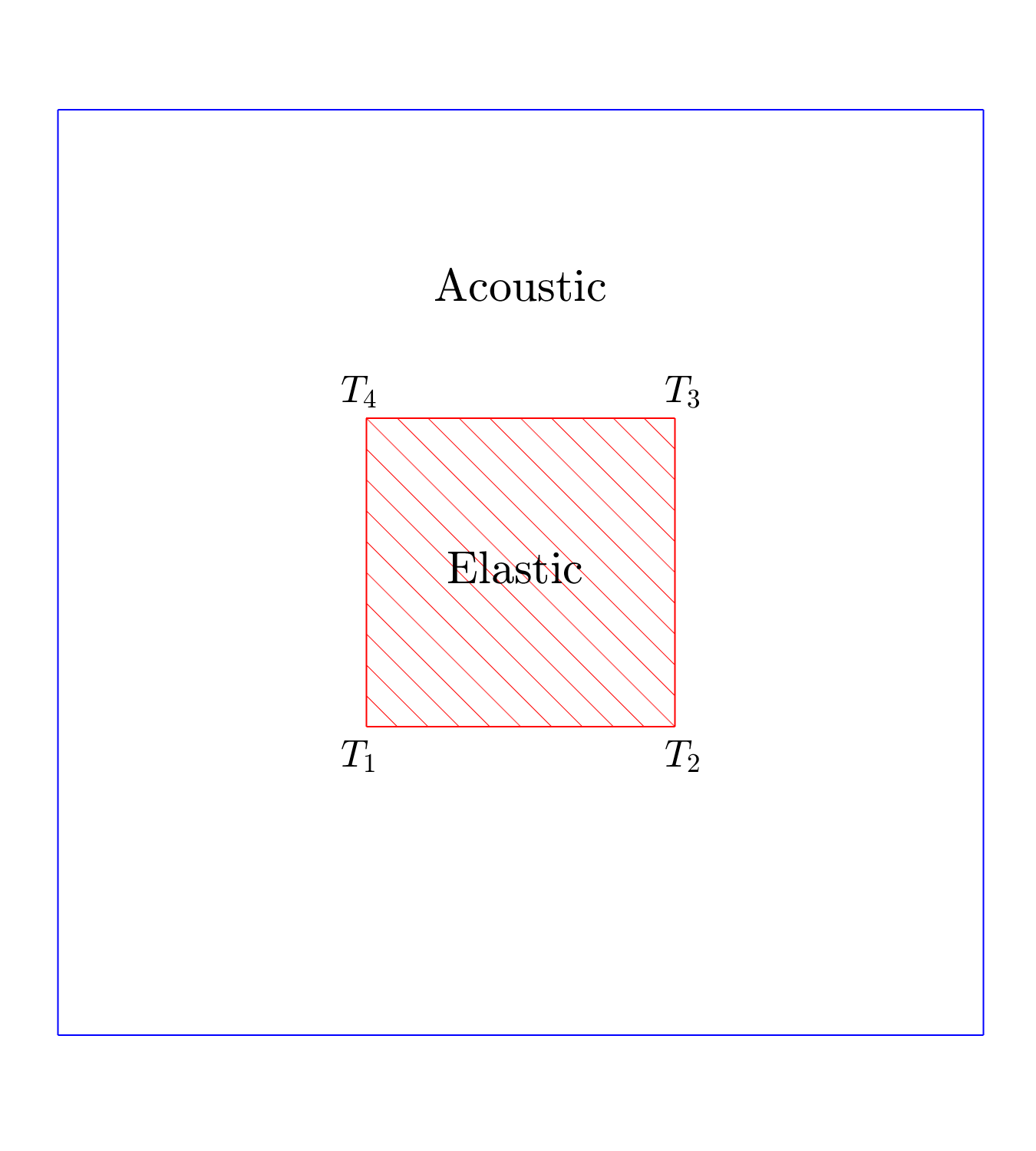}
\label{Domain_illustration}
}
\qquad
\subfigure[\!\!)]{
\captionsetup{width=1\textwidth, font=small,labelfont=small}
\centering\includegraphics[width=0.45\textwidth, scale=0.1275]{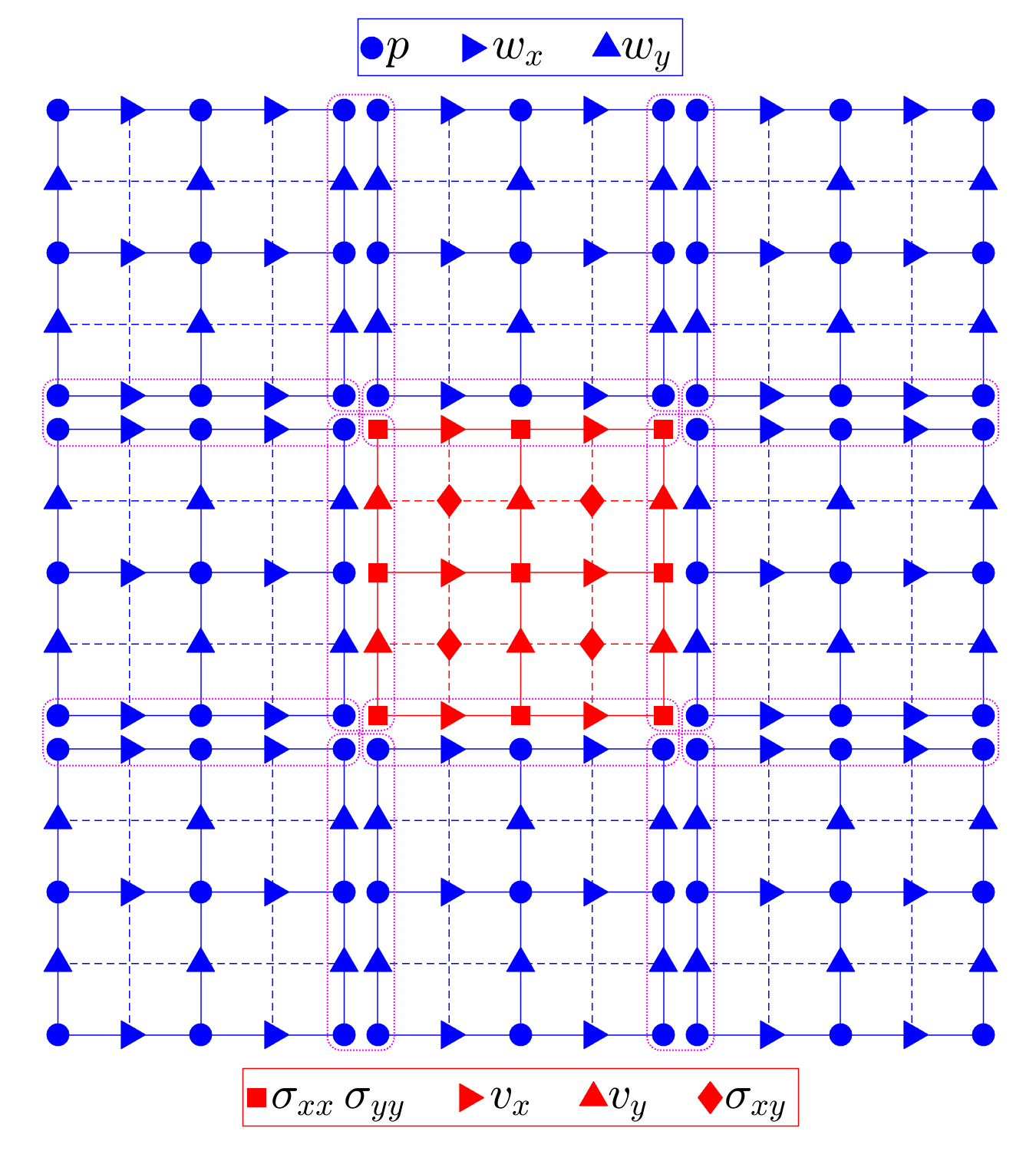}
\label{Grid_configuration}
}
%
\caption{Illustration of the geometric configuration of the simulation domain and the associated discretization grids. 
(a) The target (elastic) region is enclosed in the background (acoustic) region;
(b) Staggered grids are used to discretize the wave systems in both regions. Interfaces are encircled by the magenta lines.
}
\label{Domains_and_Grids}
\end{figure}

\textcolor{black}{
Figure \ref{Domain_illustration}} illustrates the abstracted geometric configuration under consideration, where a rectangular target region is enclosed in the background region.
An isotropic elastic medium is considered for the target region while an acoustic medium is considered for the background region. 
The respective equation systems describing the wave propagation in these two regions are presented in the subsequent sections.

For spatial discretization, we adopt the staggered grids \textcolor{black}{finite-difference} approach that dates back to \citet{yee1966numerical} and has enjoyed its popularity in seismic exploration studies; see \citet{madariaga1976dynamics}, \citet{virieux1984sh}, \citet{virieux1986p}, \citet{levander1988fourth}, among others.
The staggered discretization grids are illustrated in \textcolor{black}{Figure \ref{Grid_configuration}}.
Interfaces between the two regions are duplicated and are included in the discretizations of both sides. 

We note here that in \textcolor{black}{Figure \ref{Grid_configuration}}, the entire 2D simulation domain is divided into nine rectangular blocks. 
Artificial interfaces are drawn between adjacent acoustic blocks as well. 
These artificial interfaces are also addressed using the SBP-SAT technique.
Interested readers may consult \citet{gao2019sbp} for the SATs associated with these artificial interfaces.
Discretizations over the entire simulation domain are then addressed in a block-wise manner, which eases the upcoming discussion. 
The actual implementation is also organized in such a block-wise manner, which modularizes the code nicely.

\subsection{Acoustic background region} \label{Section_background}
For the background region, i.e., the entire simulation domain except the target region as illustrated in \textcolor{black}{Figure \ref{Domain_illustration}}, the wave propagation is described by the 2D acoustic wave system posed in the following first-order form:
\begin{subequations}
\label{Acoustic_wave_equation_PDT_2D_analysis}
\begin{empheq}[left=\empheqlbrace]{alignat = 2}
\displaystyle \rho \frac{\partial w_x}{\partial t} \enskip &= \enskip \displaystyle \! \frac{\partial p}{\partial x}; 
\label{Acoustic_wave_equation_PDT_2D_analysis_Wx} \\
\displaystyle \rho \frac{\partial w_y}{\partial t} \enskip &= \enskip \displaystyle \! \frac{\partial p}{\partial y}; 
\label{Acoustic_wave_equation_PDT_2D_analysis_Wy} \\
\displaystyle \frac{1}{\rho c^2} \frac{\partial p}{\partial t} \enskip &= \enskip \displaystyle \! \frac{\partial w_x}{\partial x} + \frac{\partial w_y}{\partial y},
\label{Acoustic_wave_equation_PDT_2D_analysis_P}
\end{empheq}
\end{subequations}
where physical parameters $\rho$ and $c$ represent density and compressional wave-speed, respectively; solution variables $w_x$, $w_y$, and $p$ represent horizontal particle velocity, vertical particle velocity, and the negative of pressure, respectively.
The choice of expressing the acoustic wave equation in terms of the negative of pressure, as opposed to pressure itself, improves the consistency between the above acoustic wave system and the elastic wave system presented later, which in turn improves the clarity and consistency of the interface treatment presented \textcolor{black}{below}.

The kinetic and potential energy density functions associated with equation \ref{Acoustic_wave_equation_PDT_2D_analysis} are defined as
\begin{equation}
\label{Acoustic_continuous_energy_density}
\varrho_k^A = \tfrac{1}{2} \rho w_i w_i \text{\quad and \quad} \varrho_p^A = \tfrac{1}{2} \tfrac{1}{\rho c^2} p^2,
\end{equation}
respectively, where the Einstein summation convention applies to the repeated subscript index $i$.
The total continuous energy associated with equation \ref{Acoustic_wave_equation_PDT_2D_analysis} is then defined as 
\begin{equation}
\label{Acoustic_continuous_energy}
e^A = \int_{\Omega_B} \tfrac{1}{2} \rho w_i w_i \ d_{\Omega_B} + \int_{\Omega_B}  \tfrac{1}{2} \tfrac{1}{\rho c^2} p^2 \, d_{\Omega_B},
\end{equation}
where $\Omega_B$ stands for the background region. 
Differentiating equation \ref{Acoustic_continuous_energy} with respect to time, substituting in equations \ref{Acoustic_wave_equation_PDT_2D_analysis_Wx} - \ref{Acoustic_wave_equation_PDT_2D_analysis_P}, and then applying the divergence theorem, we arrive at
\begin{equation}
\label{Acoustic_continuous_energy_analysis}
\frac{d e^A}{d t} = \int_{\partial \Omega_B} p w_i n_i \ d_{\partial \Omega_B}.
\end{equation}
In equation \ref{Acoustic_continuous_energy_analysis}, the Einstein summation convention applies to subscript index $i$; $n_i$ represents the $i$th component of the outward normal direction vector; $\partial \Omega_B$ denotes the boundaries of $\Omega_B$.  
We note here that $\partial \Omega_B$ includes both the exterior boundaries and the interfaces shared with the target region.
Assuming that \textcolor{black}{free-surface} boundary condition, i.e., $p = 0$, is associated with the exterior boundaries, we have that $\tfrac{d e^A}{d t}$ in equation \ref{Acoustic_continuous_energy_analysis} reduces to the interfaces only.

\subsection{Elastic target region}\label{Section_target}
For the target region illustrated in \textcolor{black}{Figure \ref{Domain_illustration}}, the wave propagation is described by the 2D isotropic elastic wave system posed in the following first-order 
form: 
\begin{subequations}
\label{Elastic_wave_equation_PDT_2D_analysis}
\begin{empheq}[left=\empheqlbrace]{alignat = 2}
\displaystyle \rho \frac{\partial v_x}{\partial t} \enskip &= \enskip \displaystyle \frac{\partial \sigma_{xx}}{\partial x} + \frac{\partial \sigma_{xy}}{\partial y}; \label{Elastic_wave_equation_PDT_2D_analysis_Vx} \\
\displaystyle \rho \frac{\partial v_y}{\partial t} \enskip &= \enskip \displaystyle \frac{\partial \sigma_{xy}}{\partial x} + \frac{\partial \sigma_{yy}}{\partial y}; 
\label{Elastic_wave_equation_PDT_2D_analysis_Vy} \\
\displaystyle s_{xxkl} \frac{\partial \sigma_{kl}}{\partial t} \enskip &= \enskip \displaystyle \frac{\partial v_x}{\partial x}; 
\label{Elastic_wave_equation_PDT_2D_analysis_Sxx} \\
\displaystyle s_{xykl} \frac{\partial \sigma_{kl}}{\partial t} \enskip &= \enskip \displaystyle \frac{1}{2} \left( \frac{\partial v_y}{\partial x} + \frac{\partial v_x}{\partial y} \right); 
\label{Elastic_wave_equation_PDT_2D_analysis_Sxy} \\
\displaystyle s_{yykl} \frac{\partial \sigma_{kl}}{\partial t} \enskip &= \enskip \displaystyle \frac{\partial v_y}{\partial y},
\label{Elastic_wave_equation_PDT_2D_analysis_Syy}
\end{empheq}
\end{subequations}
where solution variables $v_x$ and $v_y$ represent horizontal and vertical particle velocities, respectively, while $\sigma_{xx}$, $\sigma_{xy}$, and $\sigma_{yy}$ are components of the stress tensor.
Physical parameters $s_{xxkl}$, $s_{xykl}$ and $s_{yykl}$ are components of the compliance tensor, which can be expressed in terms of the Lam\'{e} parameters $\lambda$ and $\mu$ for isotropic media.  
However, their exact expressions are not needed for the upcoming discussion, and are omitted here.
In equations \ref{Elastic_wave_equation_PDT_2D_analysis_Sxx} -  \ref{Elastic_wave_equation_PDT_2D_analysis_Syy}, the Einstein summation convention applies to subscript indices $k$ and $l$.

The kinetic and potential energy density functions associated with equation \ref{Elastic_wave_equation_PDT_2D_analysis} are defined as 
\begin{equation}
\label{Elastic_continuous_energy_density}
\varrho_k^E = \tfrac{1}{2} \rho v_i v_i  \text{\quad and \quad}  
\varrho_p^E = \tfrac{1}{2} \sigma_{ij} s_{ijkl} \sigma_{kl}\, ,
\end{equation}
respectively, where the Einstein summation convention applies to subscript indices $i$, $j$, $k$ and $l$.
The total continuous energy associated with equation \ref{Elastic_wave_equation_PDT_2D_analysis} is then defined as
\begin{equation}
\label{Elastic_continuous_energy}
e^E = \int_{\Omega_T} \tfrac{1}{2} \rho v_i v_i \ d_{\Omega_T} + \int_{\Omega_T} \tfrac{1}{2} \sigma_{ij} s_{ijkl} \sigma_{kl} \ d_{\Omega_T},
\end{equation}
where $\Omega_T$ stands for the target region. 
Similar to the acoustic case, after differentiating equation \ref{Elastic_continuous_energy} with respect to time, substituting in equations \ref{Elastic_wave_equation_PDT_2D_analysis_Vx} - \ref{Elastic_wave_equation_PDT_2D_analysis_Syy}, and then applying the divergence theorem, we arrive at
\begin{equation}
\label{Elastic_continuous_energy_analysis}
\frac{d e^E}{d t} = \int_{\partial \Omega_T} v_i \sigma_{ij} n_j \ d_{\partial \Omega_T}\, .
\end{equation}
In equation \ref{Elastic_continuous_energy_analysis}, the Einstein summation convention applies to subscript indices $i$ and $j$; $n_j$ represents the $j$th component of the outward normal direction vector; $\partial \Omega_T$ denotes the boundaries of $\Omega_T$, i.e., the interfaces. 
We observe from equation \ref{Elastic_continuous_energy_analysis} that $\frac{d e^E}{d t}$ reduces to the interfaces only.

\subsection{\textcolor{black}{A remark on the forms of the wave systems}}
\textcolor{black}{
We note here that in both the acoustic wave system \textcolor{black}{(equations \ref{Acoustic_wave_equation_PDT_2D_analysis})} and the elastic wave system \textcolor{black}{(equations \ref{Elastic_wave_equation_PDT_2D_analysis})}, the physical parameters (i.e., $\rho$, $c$, and $s_{ijkl}$) appear on the left hand side together with the temporal derivatives, 
in contrast to their equivalent forms more commonly used in the literature (cf. equations \ref{Acoustic_wave_equation_PDT_2D_implementation} and \ref{Elastic_wave_equation_PDT_2D_implementation}), where the physical parameters appear on the right hand side together with the spatial derivatives.}

The forms in \textcolor{black}{equations} \ref{Acoustic_wave_equation_PDT_2D_analysis} and \ref{Elastic_wave_equation_PDT_2D_analysis} are easy to maneuver in energy analysis since these physical parameters appear explicitly in the energy definitions \textcolor{black}{(equations \ref{Acoustic_continuous_energy} and \ref{Elastic_continuous_energy})}.
Following the differentiation of the energy, substitution of the wave equations becomes straightforward with \textcolor{black}{equations} \ref{Acoustic_wave_equation_PDT_2D_analysis} and \ref{Elastic_wave_equation_PDT_2D_analysis}.
Afterward, the physical parameters no longer appear in the subsequent derivations. 
This benefit of convenience is particularly valuable when deriving the modified discretized systems \textcolor{black}{(equations \ref{Modified_Acoustic_wave_equation_Discretization_2D_analysis} and \ref{Modified_Elastic_wave_equation_Discretization_2D_analysis})} that incorporate the interface conditions, since one can operate on the right hand side of the systems freely without concern for the physical parameters.

However, these forms are not convenient for implementation, particularly the elastic wave system \textcolor{black}{(equations \ref{Elastic_wave_equation_PDT_2D_analysis})}, where the stress components are tangled together by the compliance tensor. 
Their equivalent forms in \textcolor{black}{equations} \ref{Acoustic_wave_equation_PDT_2D_implementation} and \ref{Elastic_wave_equation_PDT_2D_implementation}
are more suited for this purpose.
Fortunately, as explained in \textcolor{black}{Appendix \ref{Remark_implementation}}, the interface treatment derived based on the systems \textcolor{black}{in equations} \ref{Acoustic_wave_equation_PDT_2D_analysis} and \ref{Elastic_wave_equation_PDT_2D_analysis} can be easily translated to their equivalent forms by viewing the appended penalty terms as modifications to the corresponding spatial derivative approximations. 
Usage of the less common forms presented above is merely a mathematical device to ease the effort for derivation and poses no hindrance for a standard and efficient implementation.

\section{Spatial Discretization} \label{section_discretization}
To discretize the spatial derivatives appearing in equations \ref{Acoustic_wave_equation_PDT_2D_analysis} and \ref{Elastic_wave_equation_PDT_2D_analysis}, we use the SBP \textcolor{black}{finite-difference} operators, which are variants of the standard \textcolor{black}{finite-difference} operators with special adaptations to boundaries or interfaces so that they mimic the integration-by-parts property of the corresponding continuous operators.
For acoustic and elastic wave systems, usage of such SBP operators leads to well-defined discrete energies that mimic the behaviors of the continuous energies as illustrated in equations \ref{Acoustic_continuous_energy_analysis} and \ref{Elastic_continuous_energy_analysis}, i.e., time derivatives of the continuous energies reduce to the domain boundaries only. 

Moreover, SBP \textcolor{black}{finite-difference} operators themselves usually do not concern boundary or interface conditions, which are typically addressed by appending penalty terms, i.e., SATs, to the semi-discretized systems.
When combined with properly designed SATs, the SBP-SAT technique can deliver stable and accurate semi-discretizations with mathematical guarantee.
For an illustration of the SBP-SAT technique applied on staggered grids for seismic wave problems, interested readers may consult \textcolor{black}{\citet{gao2019sbp}} and \textcolor{black}{\citet{o2017energy}}. 
\textcolor{black}{A supplementary material is also provided to demonstrate these concepts in the context related to this article.}

Previous works on wave propagation using the SBP-SAT technique can be found in \textcolor{black}{\cite{appelo2009stable}, \cite{petersson2015wave}, \cite{wang2016high}, \cite{o2017energy}, \cite{gao2019sbp}, among others.} 
For this study, we adopt the staggered grid SBP \textcolor{black}{finite-difference} operators presented in \textcolor{black}{\citet{gao2019sbp}}, which are included in Appendix~\ref{Appendix_1D_SBP_operators} to make this work self-contained.
However, the interface treatment presented here can be easily adjusted for other choices of SBP operators.
The focus of this work is rather on the proper penalty terms, i.e., SATs, that couple the elastic and acoustic discretizations together.

\begin{figure}[H]
\captionsetup{width=1\textwidth, font=small,labelfont=small}
\centering\includegraphics[scale=0.2]{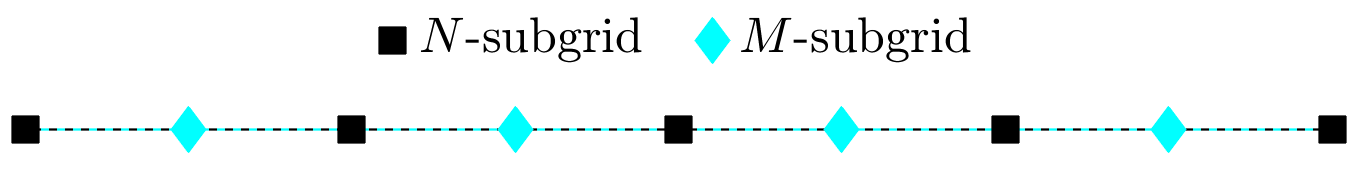}
\vspace{-0.5em}
\caption{Illustration of the 1D subgrids. 
The $N$-subgrid aligns with both boundaries; the $M$-subgrid is staggered for half the grid spacing from the $N$-subgrid and has one less grid point.}
\label{N_and_M_grids_1D}
\end{figure}

Specifically, the 1D SBP operators used here are associated with the staggered grids illustrated in Figure \ref{N_and_M_grids_1D}, which include the 1D SBP \textcolor{black}{finite-difference} operators, denoted as $\mathcal D^N$ and $\mathcal D^M$, and their associated 1D norm matrices, denoted as $\mathcal A^M$ and $\mathcal A^N$, respectively. 
Their exact forms can be found in equation \ref{SBP_matrices_1D}.
Superscripts $^N$ and $^M$ are appended to indicate the subgrids \textcolor{black}{on which} these operators act. 
For example, $\mathcal D^N$ can be applied to a discrete vector defined on the $N$-subgrid.  
The 2D SBP operators introduced later are constructed from these 1D SBP operators via tensor product.

In the interior, $\mathcal D^N$ and $\mathcal D^M$ are identical to the \textcolor{black}{finite-difference} operators presented in \cite{levander1988fourth} and are characterized by the \textcolor{black}{fourth-order} stencil $\nicefrac{ [\nicefrac{1}{24}, \enskip -\nicefrac{9}{8}, \enskip \nicefrac{9}{8}, \enskip -\nicefrac{1}{24}] }{ \Delta x }$. 
Near the boundaries, the stencils in $\mathcal D^N$ and $\mathcal D^M$ are adapted as illustrated in equations \ref{SBP_matrices_1D_DP} and \ref{SBP_matrices_1D_DV}, respectively, whose approximation order reduces to the \textcolor{black}{second order}.
The adapted boundary region includes four $N$-subgrid points and three $M$-subgrid points.

The associated norm matrices enter the definitions of the discrete physical energies as the counterparts of the integrals from the continuous energies.
Interested readers may refer to \cite{hicken2013summation} for more information on the connection between norm matrices and quadrature rules.
Although these norm matrices will not show up explicitly in implementation, they will play a pivotal role in deriving the proper interface treatment.

In this study, we limit our discussion to the case of diagonal norm matrices, i.e., $\mathcal A^M$ and $\mathcal A^N$ are diagonal, and so are the 2D norm matrices constructed from them as shown later in equations \ref{Acoustic_2D_norm_matrices} and \ref{Elastic_2D_norm_matrices}. 
Moreover, the coefficient matrices, e.g., $\boldsymbol{\rho}^{W_x}$ in equation \ref{Acoustic_wave_equation_Discretization_2D_analysis_W_x}, are also diagonal with the \textcolor{black}{finite-difference} discretization approach. 
Consequently, the two sets of matrices are symmetric and commutable, provided that the matrix sizes are the same, which makes equations \ref{Acoustic_discrete_energy} and \ref{Elastic_discrete_energy} proper definitions for discrete energies.

By design, the aforementioned 1D operators satisfy the following SBP property
\begin{equation}
\label{1D_SBP_property}
\mathcal A^N \mathcal D^M + \left( \mathcal A^M \mathcal D^N\right)^T 
=
\mathcal E^R \left(\mathcal P^R\right)^T \! - \, \mathcal E^L \left(\mathcal P^L\right)^T\!,
\end{equation}
where $\mathcal E^L$ and $\mathcal E^R$ are the canonical basis vectors that select the values of the solution vectors defined on the $N$-subgrid at the left and right endpoints, respectively; $\mathcal P^L$ and $\mathcal P^R$ are projection vectors that extrapolate the values of the solution vectors defined on the $M$-subgrid to the left and right endpoints, respectively.
The exact forms of $\mathcal E^L$, $\mathcal E^R$, $\mathcal P^L$, and $\mathcal P^R$ are included in equation \ref{Vectors_E_and_P}.

With the notations and definitions introduced above,  
we are now ready to discuss the SBP-SAT discretization on the 2D simulation domain.
For the upcoming discussion, we attach subscripts $_x$ and $_y$ to the aforementioned 1D SBP operators to indicate the directions that they are associated with.

\subsection{Acoustic background region} \label{section_background_discretization}
Neglecting boundaries and interfaces for now, by replacing the spatial derivatives in the acoustic wave system \textcolor{black}{in equations} \ref{Acoustic_wave_equation_PDT_2D_analysis} with SBP \textcolor{black}{finite-difference} operators, we arrive at the following semi-discretized system on a background block:

\vspace{-1.25em}
\begin{subequations}
\label{Acoustic_wave_equation_Discretization_2D_analysis}
\begin{empheq}[left=\empheqlbrace]{alignat = 2}
\displaystyle \mathcal A^{W_x} \boldsymbol{\rho}^{W_x} \frac{d W_x}{d t} \enskip &= \enskip \displaystyle \mathcal A^{W_x} \mathcal D^P_x P; \label{Acoustic_wave_equation_Discretization_2D_analysis_W_x} \\
\displaystyle \mathcal A^{W_y} \boldsymbol{\rho}^{W_y} \frac{d W_y}{d t} \enskip &= \enskip \displaystyle \mathcal A^{W_y} \mathcal D^P_y P; 
\label{Acoustic_wave_equation_Discretization_2D_analysis_W_y} \\
\displaystyle \mathcal A^P \boldsymbol{\mathcal C}^P \frac{d P}{d t} \enskip &= \enskip \displaystyle \mathcal A^{P} \mathcal D^{W_x}_x W_x + \mathcal A^{P} \mathcal D^{W_y}_y W_y\, .
\label{Acoustic_wave_equation_Discretization_2D_analysis_P}
\end{empheq}
\end{subequations}
In equation \ref{Acoustic_wave_equation_Discretization_2D_analysis}, $W_x$, $W_y$, and $P$ are discrete solution vectors corresponding to solution variables $w_x$, $w_y$, and $p$, respectively; 
$\boldsymbol{\rho}^{W_x}$, $\boldsymbol{\rho}^{W_y}$, and $\boldsymbol{\mathcal C}^P$ are diagonal matrices containing the discrete physical parameters on the respective subgrids, with $\boldsymbol{\rho}^{W_x}$ and $\boldsymbol{\rho}^{W_y}$ corresponding to density $\rho$ and $\boldsymbol{\mathcal C}^P$ corresponding to compressibility $\frac{1}{\rho c^2}$; 
$\mathcal D^P_x$, $\mathcal D^P_y$, $\mathcal D^{W_x}_x$, and $\mathcal D^{W_y}_y$ are the 2D SBP \textcolor{black}{finite-difference} operators whose superscripts indicate the solution variables that they act on and subscripts indicate the directions of differentiation; 
finally, $\mathcal A^{W_x}$, $\mathcal A^{W_y}$, and $\mathcal A^{P}$ are the 2D norm matrices, which are redundant in equation \ref{Acoustic_wave_equation_Discretization_2D_analysis}, 
but will play an important role in the upcoming discrete energy analysis and in derivation of the proper interface treatment.
The 2D SBP operators appearing in equation \ref{Acoustic_wave_equation_Discretization_2D_analysis} are constructed from the previously introduced 1D SBP operators via tensor product.
\textcolor{black}{
Details of their construction, as well as the relationships that they satisfy, are included in Appendix \ref{Appendix_2D_SBP_operators}.
}

The discrete energy associated with the semi-discretized system \ref{Acoustic_wave_equation_Discretization_2D_analysis} is defined as
\begin{equation}
\label{Acoustic_discrete_energy}
\mathscr E^A = \tfrac{1}{2} W_i^T \left( \mathcal A^{W_i}  \boldsymbol{\rho}^{W_i} \right) W_i^{\phantom{T}} 
\!+\,
\tfrac{1}{2} P^T \left( \mathcal A^{P} \boldsymbol{\mathcal C}^{P} \right) P\, ,
\end{equation}
where the Einstein summation convention applies only to the subscript index $i$.
The above discrete energy $\mathscr E^A$ emulates the continuous energy $e^A$ defined in equation \ref{Acoustic_continuous_energy}, with the 2D norm matrices playing the role of the integrals
in equation \ref{Acoustic_continuous_energy}.

Differentiating equation \ref{Acoustic_discrete_energy} with respect to time and substituting in equations \ref{Acoustic_wave_equation_Discretization_2D_analysis_W_x} - \ref{Acoustic_wave_equation_Discretization_2D_analysis_P}, we arrive at
\begin{equation}
\normalsize
\label{Acoustic_discrete_energy_analysis}
\frac{d \mathscr E^A}{d t} = 
P^T \left[\mathcal A^P \mathcal D_x^{W_x} + \left( \mathcal A^{W_x} \mathcal D_x^P \right)^T \right] W_x
+
P^T \left[\mathcal A^P \mathcal D_y^{W_y} + \left( \mathcal A^{W_y} \mathcal D_y^P \right)^T \right] W_y\, .
\end{equation}
Further substituting in the relations from equation \ref{Acoustic_2D_SBP_relations} and applying the mixed-product property\footnotemark[5] of the tensor product operator, we arrive at
\begin{equation}
\normalsize
\label{Acoustic_discrete_energy_analysis_2}
\arraycolsep=2.0pt
\begin{array}{rcccl}
\displaystyle
\frac{d \mathscr E^A}{d t} 
&=& \displaystyle
\uline{ \left[ P^T \! \left( \mathcal E^R_x \otimes \mathcal I^N_y \right)^{\phantom{T}} \!\!\! \right]  \mathcal A^N_y  \left[ \left( \mathcal P^R_x \otimes \mathcal I^N_y \right)^T \! W_x \right] }
&-& \displaystyle 
\uwave{ \left[ P^T \! \left( \mathcal E^L_x \otimes \mathcal I^N_y \right)^{\phantom{T}} \!\!\! \right]  \mathcal A^N_y  \left[ \left( \mathcal P^L_x \otimes \mathcal I^N_y \right)^T \! W_x \right] } \\[2ex]
&+& \displaystyle
\dotuline{ \left[ P^T \! \left( \mathcal I^N_x \otimes \mathcal E^R_y \right)^{\phantom{T}} \!\!\! \right]  \mathcal A^N_x  \left[ \left( \mathcal I^N_x \otimes \mathcal P^R_y \right)^T \! W_y \right] }
&-& \displaystyle
\dashuline{ \left[ P^T \! \left( \mathcal I^N_x \otimes \mathcal E^L_y \right)^{\phantom{T}} \!\!\! \right]  \mathcal A^N_x  \left[ \left( \mathcal I^N_x \otimes \mathcal P^L_y \right)^T \! W_y \right] }.
\end{array}
\end{equation}

\footnotetext[5]{For matrices of compatible sizes, the mixed-product property states $(A \otimes B) (C \otimes D) = (AC) \otimes (BD)$.}

Recalling the effects of the selection and projection operators,  which were introduced in equation \ref{1D_SBP_property} and symbolized by $\mathcal E$ and $\mathcal P$ therein, we recognize that the quantities inside the square brackets in equation \ref{Acoustic_discrete_energy_analysis_2}, 
for example, 
\begin{equation}
\label{example_quantities_inside_brackets}
\left[ P^T \! \left( \mathcal E^R_x \otimes \mathcal I^N_y \right)^{\phantom{T}} \!\!\! \right] 
\text{\enskip and \enskip\!\!}
\left[ \left( \mathcal I^N_x \otimes \mathcal P^L_y \right)^T \! W_y \right],
\end{equation}
are approximations of the corresponding solution variables on the edges of the block region under consideration.
The specific edges are revealed by the superscripts and subscripts of the selection or projection operators involved. 
For example, the two terms in equation \ref{example_quantities_inside_brackets} approximate $p$ on the right edge and $w_y$ on the bottom edge, respectively.

With the above observation, we have that $\tfrac{d \mathscr E^A}{d t}$ in equation \ref{Acoustic_discrete_energy_analysis_2} emulates its continuous counterpart $\tfrac{d e^A}{d t}$ in equation \ref{Acoustic_continuous_energy_analysis}, with the 1D norm matrices $\mathcal A^N_x$ and $\mathcal A^N_y$ playing the role of the line integral in equation \ref{Acoustic_continuous_energy_analysis}.
More specifically, the expression of $\frac{d \mathscr E^A}{d t}$ reduces to the edges of the block region only, with the terms underlined with solid, wavy, dotted, and dashed lines in equation \ref{Acoustic_discrete_energy_analysis_2} corresponding to the right, left, top, and bottom edges, respectively.

\subsection{Elastic target region} \label{section_target_discretization}
Similar to the acoustic case discussed above, by replacing the spatial derivatives in elastic wave \textcolor{black}{equations} \ref{Elastic_wave_equation_PDT_2D_analysis} with SBP \textcolor{black}{finite-difference} operators, we arrive at the following semi-discretized system on the target region:

\vspace{-1.25em}
\begin{subequations}
\label{Elastic_wave_equation_Discretization_2D_analysis}
\begin{empheq}[left=\empheqlbrace]{alignat = 2}
\displaystyle \mathcal A^{V_x} \boldsymbol{\rho}^{V_x} \frac{d V_x}{d t} \enskip &= \enskip \displaystyle \mathcal A^{V_x} \mathcal D^{\Sigma_{xx}}_x \Sigma_{xx} + \mathcal A^{V_x} \mathcal D^{\Sigma_{xy}}_y \Sigma_{xy}; \label{Elastic_wave_equation_Discretization_2D_analysis_Vx} \\
\displaystyle \mathcal A^{V_y} \boldsymbol{\rho}^{V_y} \frac{d V_y}{d t} \enskip &= \enskip \displaystyle \mathcal A^{V_y} \mathcal D^{\Sigma_{xy}}_x \Sigma_{xy} + \mathcal A^{V_y} \mathcal D^{\Sigma_{yy}}_y \Sigma_{yy}; 
\label{Elastic_wave_equation_Discretization_2D_analysis_Vy} \\
\displaystyle \mathcal A^{\Sigma_{xx}} \mathcal S^{\Sigma_{kl}}_{xxkl} \frac{d \Sigma_{kl}}{d t} \enskip &= \enskip \displaystyle \mathcal A^{\Sigma_{xx}} \mathcal D^{V_x}_x V_x; 
\label{Elastic_wave_equation_Discretization_2D_analysis_Sxx} \\
\displaystyle \mathcal A^{\Sigma_{xy}} \mathcal S^{\Sigma_{kl}}_{xykl} \frac{d \Sigma_{kl}}{d t} \enskip &= \enskip \displaystyle \frac{1}{2} \mathcal A^{\Sigma_{xy}} \left( \mathcal D^{V_y}_x V_y + \mathcal D^{V_x}_y V_x \right); 
\label{Elastic_wave_equation_Discretization_2D_analysis_Sxy} \\
\displaystyle \mathcal A^{\Sigma_{yy}} \mathcal S^{\Sigma_{kl}}_{yykl} \frac{d \Sigma_{kl}}{d t} \enskip &= \enskip \displaystyle \mathcal A^{\Sigma_{yy}} \mathcal D^{V_y}_y V_y
\label{Elastic_wave_equation_Discretization_2D_analysis_Syy},
\end{empheq}
\end{subequations}
where the Einstein summation convention applies only to the subscript indices $k$ and $l$ in equations \ref{Elastic_wave_equation_Discretization_2D_analysis_Sxx}-\ref{Elastic_wave_equation_Discretization_2D_analysis_Syy}. 
Treatment of the elastic-acoustic interfaces, i.e., boundaries of the target region, has been neglected in the above discretization and will be addressed in the next section. 
In \textcolor{black}{equations} \ref{Elastic_wave_equation_Discretization_2D_analysis}, $V_x$ and $V_y$ are discrete solution vectors corresponding to particle velocities $v_x$ and $v_y$ from equation \ref{Elastic_wave_equation_PDT_2D_analysis}, respectively; 
$\Sigma_{kl}$ is the discrete solution vector corresponding to the stress component $\sigma_{kl}$ from equation \ref{Elastic_wave_equation_PDT_2D_analysis}; 
and finally, diagonal matrix $\mathcal S^{\Sigma_{kl}}_{ijkl}$ contains the discrete physical parameters corresponding to the compliance tensor component $s_{i j k l}$ from equation \ref{Elastic_wave_equation_PDT_2D_analysis}, with its superscript $^{\Sigma_{kl}}$ indicating the subgrid that it is associated with. 
The 2D SBP operators appearing in equation \ref{Elastic_wave_equation_Discretization_2D_analysis} have similar meanings to those from equation \ref{Acoustic_wave_equation_Discretization_2D_analysis} and are constructed in the same manner.
\textcolor{black}{
Details of their construction, as well as the relationships that they satisfy, are included in Appendix \ref{Appendix_2D_SBP_operators}.
}

The discrete energy associated with the semi-discretized system \ref{Elastic_wave_equation_Discretization_2D_analysis} is defined as
\begin{equation}
\label{Elastic_discrete_energy}
\mathscr E^E = \tfrac{1}{2} V_i^T \left( \mathcal A^{V_i}  \boldsymbol{\rho}^{V_i} \right) V_i^{\phantom{T}} 
\!+\,
\tfrac{1}{2} \Sigma_{ij}^T \left( \mathcal A^{\Sigma_{ij}}  S^{\Sigma_{kl}}_{ijkl} \right) \Sigma_{kl}^{\phantom{T}}\, ,
\end{equation}
where the Einstein summation convention applies only to the subscript indices $i$, $j$, $k$, and $l$.
The above discrete energy $\mathscr E^E$ emulates the continuous energy $e^E$ defined in equation \ref{Elastic_continuous_energy}, with the 2D norm matrices playing the role of the integrals in equation \ref{Elastic_continuous_energy}.

Differentiating equation \ref{Elastic_discrete_energy} with respect to time and then substituting in equations \ref{Elastic_wave_equation_Discretization_2D_analysis_Vx} - \ref{Elastic_wave_equation_Discretization_2D_analysis_Syy} as well as the relations from equation \ref{Elastic_2D_SBP_relations}, we eventually arrive at 
\begin{equation}
\small
\label{Elastic_discrete_energy_analysis_2}
\arraycolsep=3.0pt
\begin{array}{rclll}
\displaystyle
\frac{d \mathscr E^E}{d t} 
&=& \displaystyle
\uline{ \left[ \Sigma_{xx}^T \left( \mathcal E^R_x \otimes \mathcal I^N_y \right)^{\phantom{T}} \!\!\! \right]  \mathcal A^N_y  \left[ \left( \mathcal P^R_x \otimes \mathcal I^N_y \right)^T V_x \right] }
&-& \displaystyle 
\uwave{ \left[ \Sigma_{xx}^T \left( \mathcal E^L_x \otimes \mathcal I^N_y \right)^{\phantom{T}} \!\!\! \right]  \mathcal A^N_y  \left[ \left( \mathcal P^L_x \otimes \mathcal I^N_y \right)^T V_x \right] } \\[2ex]
&+& \displaystyle
\uline{ \left[ V_y^T \left( \mathcal E^R_x \otimes \mathcal I^M_y \right)^{\phantom{T}} \!\!\! \right]  \mathcal A^M_y  \left[ \left( \mathcal P^R_x \otimes \mathcal I^M_y \right)^T \Sigma_{xy} \right] }
&-& \displaystyle 
\uwave{ \left[ V_y^T \left( \mathcal E^L_x \otimes \mathcal I^M_y \right)^{\phantom{T}} \!\!\! \right]  \mathcal A^M_y  \left[ \left( \mathcal P^L_x \otimes \mathcal I^M_y \right)^T \Sigma_{xy} \right] } \\[2ex]
&+& \displaystyle
\dotuline{ \left[ \Sigma_{yy}^T \left( \mathcal I^N_x \otimes \mathcal E^R_y \right)^{\phantom{T}} \!\!\! \right]  \mathcal A^N_x  \left[ \left( \mathcal I^N_x \otimes \mathcal P^R_y \right)^T V_y \right] }
&-& \displaystyle
\dashuline{ \left[ \Sigma_{yy}^T \left( \mathcal I^N_x \otimes \mathcal E^L_y \right)^{\phantom{T}} \!\!\! \right]  \mathcal A^N_x  \left[ \left( \mathcal I^N_x \otimes \mathcal P^L_y \right)^T V_y \right] } \\[2ex]
&+& \displaystyle
\dotuline{ \left[ V_x^T \left( \mathcal I^M_x \otimes \mathcal E^R_y \right)^{\phantom{T}} \!\!\! \right]  \mathcal A^M_x  \left[ \left( \mathcal I^M_x \otimes \mathcal P^R_y \right)^T \Sigma_{xy} \right] }
&-& \displaystyle
\dashuline{ \left[ V_x^T \left( \mathcal I^M_x \otimes \mathcal E^L_y \right)^{\phantom{T}} \!\!\! \right]  \mathcal A^M_x  \left[ \left( \mathcal I^M_x \otimes \mathcal P^L_y \right)^T \Sigma_{xy} \right] }.
\end{array}
\end{equation}
Similar to the acoustic case discussed before, the quantities inside the square brackets in equation \ref{Elastic_discrete_energy_analysis_2} are approximations of the corresponding solution variables on the edges of the target region.

Consequently, we have that $\tfrac{d \mathscr E^E}{d t}$ in equation \ref{Elastic_discrete_energy_analysis_2} emulates its continuous counterpart $\tfrac{d e^E}{d t}$ in equation \ref{Elastic_continuous_energy_analysis}, 
with the 1D norm matrices $\mathcal A^N_x$, $\mathcal A^M_x$, $\mathcal A^N_y$, and $\mathcal A^M_y$ playing the role of the line integral in equation \ref{Elastic_continuous_energy_analysis}.
More specifically, the expression of $\frac{d \mathscr E^E}{d t}$ reduces to the edges of the target region only, with the terms underlined with solid, wavy, dotted, and dashed lines corresponding to the right, left, top, and bottom edges, respectively.

\subsection{Interface treatment} \label{section_interface_treatment}
On elastic-acoustic interfaces, we seek to impose the following interface conditions \citep[see][ p.~52]{stein2009introduction},
\begin{subequations}
\label{Interface_conditions}
\begin{alignat}{1}
\omit \hfill $w_i n_i^A + v_i n_i^E = 0$; \hfill 
\label{Interface_conditions_velocity}
\shortintertext{and}
\omit \hfill $p n_i^A + \sigma_{ij} n_j^E = 0$, \hfill
\label{Interface_conditions_stress}
\end{alignat}
\end{subequations}
where $n_i^A$ and $n_i^E$ (or $n_j^E$) represent components of the outward normal direction vectors at the interfaces from the acoustic and the elastic regions, respectively, which are the opposite of each other. 

Recalling the results from equations \ref{Acoustic_continuous_energy_analysis} and \ref{Elastic_continuous_energy_analysis}, and neglecting the exterior boundaries for now, we have that 
\textcolor{black}{
\begin{equation}
\label{Energy_conserving_continuous}
\frac{d e^A}{d t} + \frac{d e^E}{d t} 
\enskip = \enskip
\displaystyle \int_{\partial \Omega_T} p w_i n^A_i \ d_{\partial \Omega_T} \ + \ \int_{\partial \Omega_T} v_i \sigma_{ij} n^E_j \ d_{\partial \Omega_T} \enskip = \enskip 0 \, ,
\end{equation}
after first applying the interface condition \ref{Interface_conditions_stress}, then using the fact that $n_i^A$ and $n_i^E$ are the opposite of each other, and, finally, applying the interface condition \ref{Interface_conditions_velocity}.}
Equation \ref{Energy_conserving_continuous} reveals that the total continuous energy, i.e., the sum of $e^A$ 
and $e^E$, 
is conserved across the elastic-acoustic interfaces.  
As demonstrated below, this property is preserved in the discretization with the proposed interface treatment.

It suffices to illustrate the interface treatment for the left edge of the target region as shown in \textcolor{black}{Figure \ref{Domain_illustration}}, since the other edges can be addressed in a similar manner. 
This left edge is referred to as $T_1T_4$ using its two endpoints. 
For its treatment, we only need to consider the target region and the background block region left-adjacent to it, which share edge $T_1T_4$ as their interface. 

To account for the interface conditions in equation \ref{Interface_conditions} on edge $T_1T_4$, we append penalty terms, i.e., SATs, to the semi-discretized systems \ref{Acoustic_wave_equation_Discretization_2D_analysis} and \ref{Elastic_wave_equation_Discretization_2D_analysis}, leading to the following modified semi-discretized systems \ref{Modified_Acoustic_wave_equation_Discretization_2D_analysis} and \ref{Modified_Elastic_wave_equation_Discretization_2D_analysis}, respectively,

\begin{subequations}
\small
\label{Modified_Acoustic_wave_equation_Discretization_2D_analysis}
({\normalsize Acoustic})
\begin{empheq}[left=\empheqlbrace]{alignat = 2}
\label{Modified_Acoustic_wave_equation_Discretization_2D_analysis_Wx}
\displaystyle \mathcal A^{W_x} \boldsymbol{\rho}^{W_x} \frac{d W_x}{d t} 
\enskip &= \enskip \displaystyle \ \! \mathcal A^{W_x} \mathcal D^P_x P \\[-1ex] \nonumber
\enskip &+ \enskip \eta_A^{W_x}  \left[ \mathcal P^R_x \otimes \mathcal I^N_y \right]  \mathcal A^N_y  \left\{ \left[ \left( \mathcal E^R_x \right)^T \!\! \otimes \mathcal I^N_y\right] P - \left[ \left( \mathcal E^L_x \right)^T \!\! \otimes \mathcal I^N_y\right] \Sigma_{xx} \right\}; \\
\label{Modified_Acoustic_wave_equation_Discretization_2D_analysis_Wy}
\displaystyle \mathcal A^{W_y} \boldsymbol{\rho}^{W_y} \frac{d W_y}{d t} 
\enskip &= \enskip \displaystyle \ \! \mathcal A^{W_y} \mathcal D^P_y P; \\
\label{Modified_Acoustic_wave_equation_Discretization_2D_analysis_P}
\displaystyle \mathcal A^P \mathcal C^P \frac{d P}{d t} 
\enskip &= \enskip \displaystyle \ \! \mathcal A^{P} \mathcal D^{W_x}_x W_x + \mathcal A^{P} \mathcal D^{W_y}_y W_y \\[-1ex] \nonumber
\enskip &+ \enskip \eta_A^P  \left[ \mathcal E^R_x \otimes \mathcal I^N_y \right]  \mathcal A^N_y  \left\{ \left[ \left( \mathcal P^R_x \right)^T \!\! \otimes \mathcal I^N_y\right] W_x - \left[ \left( \mathcal P^L_x \right)^T \!\! \otimes \mathcal I^N_y\right] V_x \right\}, 
\end{empheq}
\end{subequations}

\begin{subequations}
\small
\label{Modified_Elastic_wave_equation_Discretization_2D_analysis}
({\normalsize Elastic})
\begin{empheq}[left=\empheqlbrace]{alignat = 2}
\label{Modified_Elastic_wave_equation_Discretization_2D_analysis_Vx}
\displaystyle \mathcal A^{V_x} \boldsymbol{\rho}^{V_x} \frac{d V_x}{d t} 
\enskip &= \enskip \displaystyle \mathcal A^{V_x} \mathcal D^{\Sigma_{xx}}_x \Sigma_{xx} + \mathcal A^{V_x} \mathcal D^{\Sigma_{xy}}_y \Sigma_{xy} \\[-1ex] \nonumber
\enskip &+ \enskip \eta_E^{V_x}  \left[ \mathcal P^L_x \otimes \mathcal I^N_y \right]  \mathcal A^N_y  \left\{ \left[ \left( \mathcal E^L_x \right)^T \!\! \otimes \mathcal I^N_y\right] \Sigma_{xx} - \left[ \left( \mathcal E^R_x \right)^T \!\! \otimes \mathcal I^N_y\right] P \right\}; \\
\label{Modified_Elastic_wave_equation_Discretization_2D_analysis_Vy}
\displaystyle \mathcal A^{V_y} \boldsymbol{\rho}^{V_y} \frac{d V_y}{d t} 
\enskip &= \enskip \displaystyle \mathcal A^{V_y} \mathcal D^{\Sigma_{xy}}_x \Sigma_{xy} + \mathcal A^{V_y} \mathcal D^{\Sigma_{yy}}_y \Sigma_{yy} \\[-1ex] \nonumber
\enskip &+ \enskip \eta_E^{V_y}  \left[ \mathcal E^L_x \otimes \mathcal I^M_y \right]  \mathcal A^M_y  \left\{ \left[ \left( \mathcal P^L_x \right)^T \!\! \otimes \mathcal I^M_y\right] \Sigma_{xy} - \boldsymbol 0 \right\}; \\
\label{Modified_Elastic_wave_equation_Discretization_2D_analysis_Sxx}
\displaystyle \mathcal A^{\Sigma_{xx}} \mathcal S^{\Sigma_{kl}}_{xxkl} \frac{d \Sigma_{kl}}{d t} 
\enskip &= \enskip \displaystyle \mathcal A^{\Sigma_{xx}} \mathcal D^{V_x}_x V_x \\[-1ex] \nonumber
\enskip &+ \enskip \eta_E^{\Sigma_{xx}}  \left[ \mathcal E^L_x \otimes \mathcal I^N_y \right]  \mathcal A^N_y  \left\{ \left[ \left( \mathcal P^L_x \right)^T \!\! \otimes \mathcal I^N_y\right] V_x - \left[ \left( \mathcal P^R_x \right)^T \!\! \otimes \mathcal I^N_y\right] W_x \right\}; \\
\label{Modified_Elastic_wave_equation_Discretization_2D_analysis_Sxy}
\displaystyle \mathcal A^{\Sigma_{xy}} \mathcal S^{\Sigma_{kl}}_{xykl} \frac{d \Sigma_{kl}}{d t} 
\enskip &= \enskip \displaystyle \tfrac{1}{2} \mathcal A^{\Sigma_{xy}} \left( \mathcal D^{V_y}_x V_y + \mathcal D^{V_x}_y V_x \right); \\
\label{Modified_Elastic_wave_equation_Discretization_2D_analysis_Syy}
\displaystyle \mathcal A^{\Sigma_{yy}} \mathcal S^{\Sigma_{kl}}_{yykl} \frac{d \Sigma_{kl}}{d t} 
\enskip &= \enskip \displaystyle \mathcal A^{\Sigma_{yy}} \mathcal D^{V_y}_y V_y.
\end{empheq}
\end{subequations}
We note here that the modified system \textcolor{black}{of equations} \ref{Modified_Acoustic_wave_equation_Discretization_2D_analysis} applies only to the background block region left-adjacent to the target region. 
In these two modified semi-discretized systems, 
the penalty terms in equations \ref{Modified_Acoustic_wave_equation_Discretization_2D_analysis_P} and \ref{Modified_Elastic_wave_equation_Discretization_2D_analysis_Sxx} account for the interface condition \textcolor{black}{stated in equation \ref{Interface_conditions_velocity}}, 
while the penalty terms in equations \ref{Modified_Acoustic_wave_equation_Discretization_2D_analysis_Wx} and \ref{Modified_Elastic_wave_equation_Discretization_2D_analysis_Vx}, as well as \textcolor{black}{equation} 
\ref{Modified_Elastic_wave_equation_Discretization_2D_analysis_Vy}, account for the interface condition \textcolor{black}{stated in equation \ref{Interface_conditions_stress}}.

The scalars $\eta_A^{W_x}$ and $\eta_A^P$ from equation \ref{Modified_Acoustic_wave_equation_Discretization_2D_analysis}, as well as $\eta_E^{V_x}$, $\eta_E^{V_y}$ and $\eta_E^{\Sigma_{xx}}$ from equation \ref{Modified_Elastic_wave_equation_Discretization_2D_analysis}, are penalty parameters.
By setting them to be
\begin{equation}
\label{value_penalty_parameters}
\eta_A^{W_x} = \eta_A^P = - \nicefrac{1}{2}, \enskip
\eta_E^{V_x} = \eta_E^{\Sigma_{xx}} = \nicefrac{1}{2}, \enskip \text{and} \enskip 
\eta_E^{V_y} = 1,
\end{equation}
the energy-conserving property across edge $T_1T_4$ is preserved in the semi-discretization.

To verify this, we differentiate the sum of $\mathscr E^A$ and $\mathscr E^E$ with respect to time, similarly to what we did before in equations \ref{Acoustic_discrete_energy_analysis_2} and \ref{Elastic_discrete_energy_analysis_2}.
However, this time, we only keep track of those terms that are associated with edge $T_1T_4$. 
After substituting in the equations from \ref{Modified_Acoustic_wave_equation_Discretization_2D_analysis} and \ref{Modified_Elastic_wave_equation_Discretization_2D_analysis}, we arrive at 

\vspace{-1.5em}
\begin{subequations}
\footnotesize
\label{Discrete_energy_analysis_T1T4}
\begin{alignat}{5}
& \left( \frac{d \mathscr E^A}{d t} + \frac{d \mathscr E^E}{d t} \right) {\Bigg |}_{T_1T_4} & \nonumber \\
= \ &
\quad\enskip\enskip \uline{ \left[ P^T\!\! \left( \mathcal E^R_x \otimes \mathcal I^N_y \right)^{\phantom{T}} \!\!\! \right]  \mathcal A^N_y  \left[ \left( \mathcal P^R_x \otimes \mathcal I^N_y \right)^T \! W_x \right] } \label{Discrete_energy_analysis_T1T4_a} \\
- \ & 
\quad\enskip\enskip \uwave{ \left[ \Sigma_{xx}^T \left( \mathcal E^L_x \otimes \mathcal I^N_y \right)^{\phantom{T}} \!\!\! \right]  \mathcal A^N_y  \left[ \left( \mathcal P^L_x \otimes \mathcal I^N_y \right)^T \! V_x \right] }
& \ - \ &
\quad\enskip\enskip \uuline{ \left[ V_y^T \left( \mathcal E^L_x \otimes \mathcal I^M_y \right)^{\phantom{T}} \!\!\! \right]  \mathcal A^M_y  \left[ \left( \mathcal P^L_x \otimes \mathcal I^M_y \right)^T \! \Sigma_{xy} \right] } 
\label{Discrete_energy_analysis_T1T4_b} \\
+ \ & \eta_A^{W_x}  \uline{ \left[ W_x^T \left(\mathcal P^R_x \otimes \mathcal I^N_y\right)^{\phantom{T}} \!\!\! \right]  \mathcal A^N_y  \left[ \left( \mathcal E^R_x \otimes \mathcal I^N_y \right)^T \! P \right] }
& \ - \ &
\eta_A^{W_x}  \dotuline{ \left[ W_x^T \left(\mathcal P^R_x \otimes \mathcal I^N_y\right)^{\phantom{T}} \!\!\! \right]  \mathcal A^N_y  \left[ \left( \mathcal E^L_x \otimes \mathcal I^N_y \right)^T \! \Sigma_{xx} \right] } \label{Discrete_energy_analysis_T1T4_c} \\
+ \ & \eta_A^P \enskip \uline{ \left[ P^T\!\! \left( \mathcal E^R_x \otimes \mathcal I^N_y \right)^{\phantom{T}} \!\!\! \right]  \mathcal A^N_y  \left[ \left( \mathcal P^R_x \otimes \mathcal I^N_y \right)^T \! W_x \right] }
& \ - \ &
\eta_A^P \enskip \dashuline{ \left[ P^T\!\! \left( \mathcal E^R_x \otimes \mathcal I^N_y \right)^{\phantom{T}} \!\!\! \right]  \mathcal A^N_y  \left[ \left( \mathcal P^L_x \otimes \mathcal I^N_y \right)^T \! V_x \right] } \label{Discrete_energy_analysis_T1T4_d} \\
+ \ & \eta_E^{V_x}  \uwave{ \left[ V_x^T \left(\mathcal P^L_x \otimes \mathcal I^N_y\right)^{\phantom{T}} \!\!\! \right]  \mathcal A^N_y  \left[ \left( \mathcal E^L_x \otimes \mathcal I^N_y \right)^T \! \Sigma_{xx} \right] }
& \ - \ & 
\eta_E^{V_x}  \dashuline{ \left[ V_x^T \left(\mathcal P^L_x \otimes \mathcal I^N_y\right)^{\phantom{T}} \!\!\! \right]  \mathcal A^N_y  \left[ \left( \mathcal E^R_x \otimes \mathcal I^N_y \right)^T \! P \right] } \label{Discrete_energy_analysis_T1T4_e} \\
+ \ & \eta_E^{V_y}  \uuline{ \left[ V_y^T \left(\mathcal E^L_x \otimes \mathcal I^M_y\right)^{\phantom{T}} \!\!\! \right]  \mathcal A^M_y  \left[ \left( \mathcal P^L_x \otimes \mathcal I^M_y \right)^T \! \Sigma_{xy} \right] } \label{Discrete_energy_analysis_T1T4_f} \\
+ \ & 
\eta_E^{\Sigma_{xx}}  \uwave{ \left[ \Sigma_{xx}^T \left(\mathcal E^L_x \otimes \mathcal I^N_y\right)^{\phantom{T}} \!\!\! \right]  \mathcal A^N_y  \left[ \left( \mathcal P^L_x \otimes \mathcal I^N_y \right)^T \! V_x \right] }
& \ - \ & 
\eta_E^{\Sigma_{xx}}  \dotuline{ \left[ \Sigma_{xx}^T \left(\mathcal E^L_x \otimes \mathcal I^N_y\right)^{\phantom{T}} \!\!\! \right]  \mathcal A^N_y \left[ \left( \mathcal P^R_x \otimes \mathcal I^N_y \right)^T \! W_x \right] } \label{Discrete_energy_analysis_T1T4_g},
\end{alignat}
\end{subequations}
where the term in line \ref{Discrete_energy_analysis_T1T4_a} stems from the acoustic background block region as in equation \ref{Acoustic_discrete_energy_analysis_2}; 
the terms in line \ref{Discrete_energy_analysis_T1T4_b} stem from the elastic target region as in equation \ref{Elastic_discrete_energy_analysis_2}; 
the terms in lines \ref{Discrete_energy_analysis_T1T4_c} - \ref{Discrete_energy_analysis_T1T4_g} stem from the introduced penalty terms in equations \ref{Modified_Acoustic_wave_equation_Discretization_2D_analysis_Wx},  \ref{Modified_Acoustic_wave_equation_Discretization_2D_analysis_P},  \ref{Modified_Elastic_wave_equation_Discretization_2D_analysis_Vx},
\ref{Modified_Elastic_wave_equation_Discretization_2D_analysis_Vy},
and \ref{Modified_Elastic_wave_equation_Discretization_2D_analysis_Sxx}, respectively.

In equation \ref{Discrete_energy_analysis_T1T4}, the terms underlined with the same line style are the same and therefore can be collected together.
It can be easily verified that, with the choices of penalty parameters described in equation \ref{value_penalty_parameters}, the terms underlined with the same line styles cancel each other out. 
Consequently, we arrive at 
\begin{equation}
\label{Discrete_energy_conservation_T1T4}
\left( \frac{d \mathscr E^A}{d t} + \frac{d \mathscr E^E}{d t} \right) {\Bigg |}_{T_1T_4} = \enskip 0 \, .
\end{equation}
In other words, the semi-discretization presented in equations \ref{Modified_Acoustic_wave_equation_Discretization_2D_analysis} and \ref{Modified_Elastic_wave_equation_Discretization_2D_analysis} preserves the energy-conserving property across edge $T_1T_4$. 

Similar modifications to the semi-discretized systems \textcolor{black}{presented in equations} \ref{Acoustic_wave_equation_Discretization_2D_analysis} and \ref{Elastic_wave_equation_Discretization_2D_analysis} can be made for the other three edges. 
To conserve space, these modifications are provided in the supplementary material.
Observations similar to that in equation \ref{Discrete_energy_conservation_T1T4} can be made for the other three edges as well.
After all these modifications, it can be shown that all the terms in equations \ref{Acoustic_discrete_energy_analysis_2} and \ref{Elastic_discrete_energy_analysis_2} that are associated with the elastic-acoustic interfaces get cancelled out by the introduction of the properly designed penalty terms. 

Furthermore, with proper exterior boundary conditions, e.g., the \textcolor{black}{free-surface} boundary condition $p=0$, the total continuous energy $e^A + e^E$ is conserved over the entire simulation domain, as evidenced from equations \ref{Acoustic_continuous_energy_analysis} and \ref{Energy_conserving_continuous}.  
When combined with proper treatment of these exterior boundaries as well as the acoustic-acoustic artificial interfaces; see \cite{gao2019sbp}, the overall discretization using the elastic-acoustic interface treatment described above can be shown to conserve the total discrete energy $\mathscr E^A + \mathscr E^E$ over the entire simulation domain.  
Such property will be demonstrated \textcolor{black}{below}.

As a side note, we point out here that when used to couple different parallel subdomains, the SBP-SAT technique leads to the desired features of low communication volume and simple communication pattern.
Taking the pair of equations \ref{Modified_Acoustic_wave_equation_Discretization_2D_analysis_Wx} and \ref{Modified_Elastic_wave_equation_Discretization_2D_analysis_Vx} for example, the quantity that needs to be communicated from the elastic region to the acoustic region is $\left[ \left( \mathcal E^L_x \right)^T \!\! \otimes \mathcal I^N_y\right] \Sigma_{xx}$, while $\left[ \left( \mathcal E^R_x \right)^T \!\! \otimes \mathcal I^N_y\right] P$ is the quantity that needs to be communicated reversely; both are restricted to the interface. 

\section{Numerical examples} \label{Section_numerical_examples} 
In the following, we 
demonstrate the efficacy of the presented elastic-acoustic interface treatment with numerical examples. 
For all the tests presented below, the staggered leapfrog scheme is used for time integration.
The initial status of the medium is assumed to be \textcolor{black}{quiescent}, which means that all solution components and their derivatives are zero at the beginning of the simulation.
The wave propagation is driven by a point source term that mimics the explosive source commonly used in seismic exploration surveys.

\subsection{Example 1. \textcolor{black}{Validation against analytical solution}} \label{Example_1}
In this example, we validate the presented interface treatment with the semi-analytical solution provided at \url{http://www.spice-rtn.org/library/software/EX2DELEL.html} by the SPICE project. 
A horizontal elastic-acoustic interface is considered, which separates the acoustic medium above it and the elastic medium below it. Both media are homogeneous. 
The acoustic medium is characterized by compressional wave-speed $c_p^A = 3$~m/s and density $\rho^A \!=\! 1$~kg/m$^{3}$; the elastic medium is characterized by compressional wave-speed $c_p^E = 9$~m/s, shear wave-speed $c_s^E = 5$~m/s, and density $\rho^E \!= 2$~kg/m$^{3}$. 

Both source and receiver are placed in the acoustic medium. Specifically, the source is at 0.15~m above the interface on the $P$ subgrid; the receiver is at 0.1575~m above the interface and 12~m to the right of the source on the $W_y$ subgrid. The source temporal profile is chosen as the Ricker wavelet with $5$ Hz central frequency and $0.25$~s time delay.
The maximal frequency in the source content is counted as 12.5~Hz, which, together with the minimal wave-speed, i.e., $c_p^A = 3$~m/s, leads to minimal wavelength at 0.24~m.

\begin{figure}[H]
\captionsetup{width=1\textwidth, font=small,labelfont=small}
\centering
%
\subfigure[\!\!)]{
\captionsetup{width=1\textwidth, font=small,labelfont=small}
\centering\includegraphics[width=0.475\textwidth, scale=0.2]{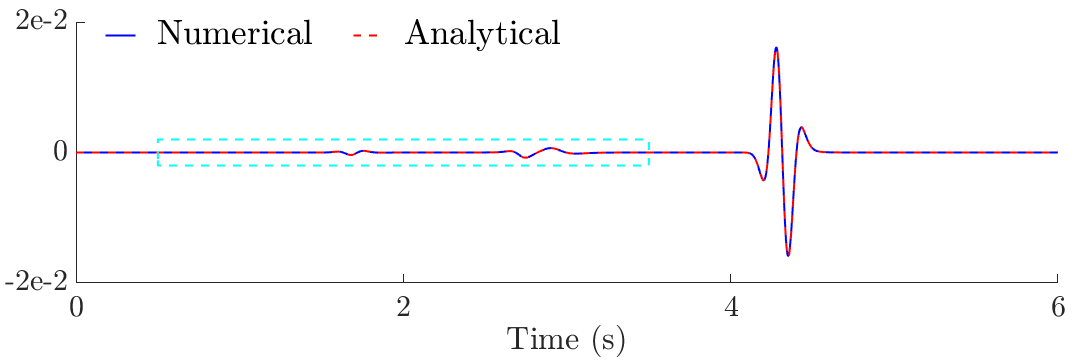}
%
}
\hfill
\subfigure[\!\!)]{
\captionsetup{width=1\textwidth, font=small,labelfont=small}
\centering\includegraphics[width=0.475\textwidth, scale=0.2]{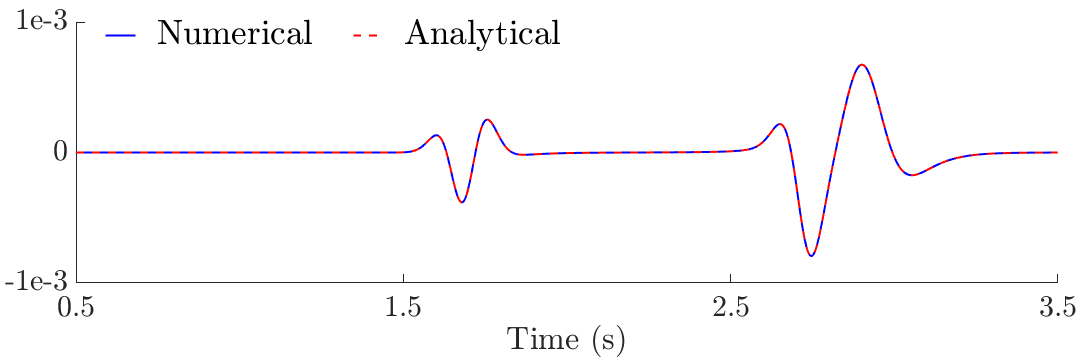}
%
}
%
\vspace{-0.5em}
\caption{Time history of $W_y$ at the receiver location from the numerical simulation compared with the semi-analytical solution. (a): the entire 6~s signal; (b): zoom-in of the segment between 0.5~s and 3.5~s.}
\label{Example_1_Seismogram}
\end{figure}

In the numerical simulation, the spatial grid spacing is chosen as $\Delta x =$ 0.015~m, which amounts to 16 grid points per minimal wavelength.
The time step length is chosen as $\Delta t =$ 8e-4~s.
Time history of $W_y$ at the receiver location from the numerical simulation is recorded for 6~s and compared with the semi-analytical solution in Figure \ref{Example_1_Seismogram}. 
We observe that the two solutions agree very well with each other.

\subsection{Example 2. \textcolor{black}{Homogeneous media}} \label{Example_2}
In this example, we consider a simulation domain as illustrated in \textcolor{black}{Figure \ref{Domain_illustration}}, where the elastic medium is surrounded by the acoustic medium. 
Both media are characterized by homogeneous parameters, which are the same as those specified in the previous example.
The entire simulation domain is divided into nine subdomains, each discretized by a set of staggered grids; see \textcolor{black}{Figure \ref{Grid_configuration}}. 
On each subdomain, there are $101 \times 101$ grid points on the $N$-subgrid, which is occupied by solution variable $p$ on an acoustic subdomain and by normal stresses $\sigma_{xx}$ and $\sigma_{yy}$ on the elastic subdomain.
An interface shared by two subdomains is duplicated and included in the discretizations of both subdomains. 

The \textcolor{black}{free-surface} boundary condition, i.e., $p=0$, is considered for all exterior boundaries, which is also imposed via the SBP-SAT technique. 
Interested readers may refer to \textcolor{black}{\citet{gao2019sbp}} for more detail.
This choice 
enables us to demonstrate the energy-conserving property of the proposed interface treatment since there is no energy loss at the \textcolor{black}{free-surface} exterior boundaries. 

A point source is placed at the center of the bottom left subdomain while a receiver is placed at the center of the top right subdomain, both on the $N$-subgrid. 
The source temporal profile is the same as in the previous example.
Time history of $P$ at the receiver location is recorded during the simulation and used later for comparison.
The spatial grid spacing is again chosen as $\Delta x =$ 0.015~m, which amounts to 16 grid points per minimal wavelength. 
The time step length is chosen as $\Delta t \approx$ 9.428e-4~s, which \textcolor{black}{corresponds} to a Courant number 0.8. 
Duration of the total simulation is set to 6~s.

\begin{figure}[H]
\captionsetup{width=1\textwidth, font=small,labelfont=small}
\centering\includegraphics[scale=0.2]{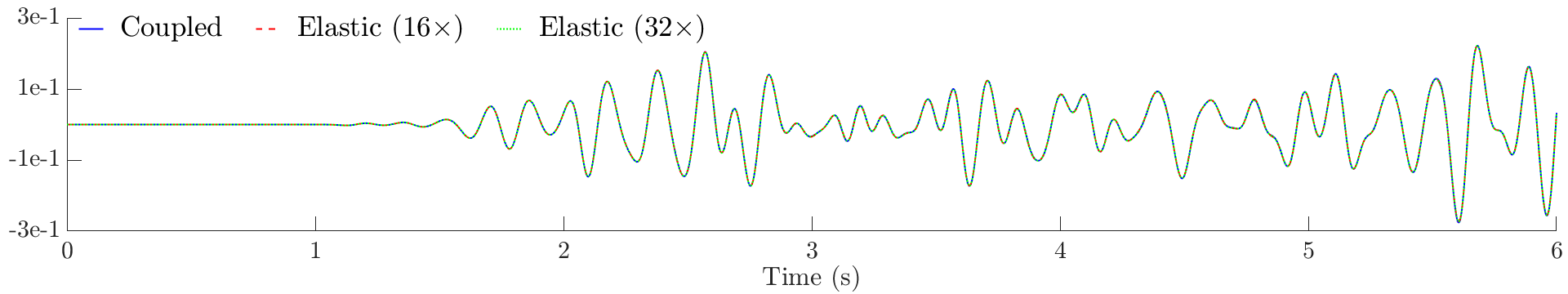}
\vspace{-.5em}
\caption{Time history of $P$ at the receiver location from the coupled simulation compared with full elastic simulation results.}
\label{Example_2_Seismogram}
\end{figure}

The recorded time history of $P$ at the receiver location from the coupled simulation is displayed in Figure \ref{Example_2_Seismogram} and labeled as `Coupled'.
For comparison, also shown in Figure \ref{Example_2_Seismogram} are the full elastic simulation results, which use the elastic wave system for the entire simulation domain and simply treat the acoustic medium in the background region as a special case of the elastic medium whose shear wave-speed value is zero.

Discretization of the elastic wave system posed in the first-order form \ref{Elastic_wave_equation_PDT_2D_implementation} can be used to stably simulate wave propagation on acoustic medium. 
In fact, one can easily verify that, by setting the shear wave-speed $c_s$ 
to zero, equation \ref{Elastic_wave_equation_PDT_2D_implementation} reduces to equation \ref{Acoustic_wave_equation_PDT_2D_implementation}.
To start, $\sigma_{xy}$ remains zero since the shear modulus $\mu = 0$;  
both $\sigma_{xx}$ and $\sigma_{yy}$ replicate $p$ since both $\lambda$ and $\lambda + 2 \mu$ reduce to the bulk modulus $\rho c^2$ in equation \ref{Acoustic_wave_equation_PDT_2D_implementation}; 
finally, with $\sigma_{xx}$ and $\sigma_{yy}$ replicating $p$ and $\sigma_{xy}$ being zero, $v_x$ and $v_y$ in equation \ref{Elastic_wave_equation_PDT_2D_implementation} reduce to $w_x$ and $w_y$ in equation \ref{Acoustic_wave_equation_PDT_2D_implementation}, respectively.

However, this implicit treatment of the elastic-acoustic interfaces will negatively affect the simulation accuracy since there is an infinite jump in the shear wave-speed value. 
Consequently, much finer grid spacing is needed to resolve the waves near the interfaces.
In Figure \ref{Example_2_Seismogram}, the full elastic simulation results using grid spacings that are $\nicefrac{1}{16}$ and $\nicefrac{1}{32}$ of that used in the coupled simulation are presented as reference solutions, which are labeled as `Elastic ($16\times$)' and `Elastic ($32\times$)', respectively.
The close proximity between these two reference solutions indicates that the waves have been well resolved in these reference simulations, even near the interfaces.

\begin{figure}[H]
\captionsetup{width=1\textwidth, font=small,labelfont=small}
\centering\includegraphics[scale=0.2]{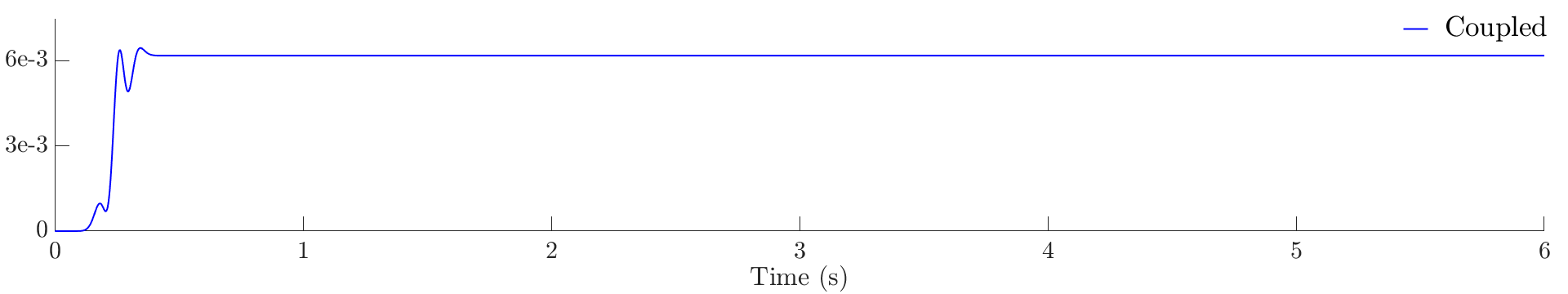}
\vspace{-.5em}
\caption{Evolution of the total discrete energy associated with the coupled simulation.}
\label{Example_2_Energy}
\end{figure}

From Figure \ref{Example_2_Seismogram}, we observe that the coupled simulation result matches very well with the reference solutions as they overlay one another. 
In addition, evolution of the total discrete energy associated with the coupled simulation, i.e., the sum of $\mathscr E^A$ from equation \ref{Acoustic_discrete_energy} over the acoustic region and $\mathscr E^E$ from equation \ref{Elastic_discrete_energy} over the elastic region, is plotted in Figure \ref{Example_2_Energy}, where we observe that the total discrete energy remains constant after the source term tapers off, which corroborates the energy-conserving property of the proposed interface treatment.

\textcolor{black}{
Since the coupled simulation and the elastic ($1\times$) simulation use the same spatial grid spacing $\Delta x$ and temporal step length $\Delta t$, 
\textcolor{black}{the} ratio of the computational costs associated with these two simulations can be estimated by the numbers of spatial derivative approximations at each grid point summed over the entire simulation domain.
This number is 4 in the acoustic region 
and 8 in the elastic region, 
which leads to an estimated ratio of $\sim\!55.6\%$ for this example.
In a serial setting, the elapsed times measured for the time loops in these two simulations are $\sim\!24.85$~s and $\sim\!41.98$~s, respectively, with a ratio of $\sim\!59.2\%$.\footnotemark[7]}
\footnotetext[7]{
\textcolor{black}{
Each simulation is run for 7 times. 
Omitting the maximum and minimum, average of the remaining 5 recorded elapsed times is reported here as the measured time.
The executables are generated with the GCC compiler with -O3 optimization flag and re-compiled after each run. 
The coupled simulation is slightly slower than the estimate predicts.
This is largely due to a suboptimal implementation of the interface treatment in our prototype code, which can be improved upon in future more performance-oriented implementations.}
}

\subsection{Example 3. \textcolor{black}{Heterogeneous media}} \label{Example_3}
\begin{figure}[H]
\captionsetup{width=1\textwidth, font=small,labelfont=small}
\centering\includegraphics[scale=0.15]{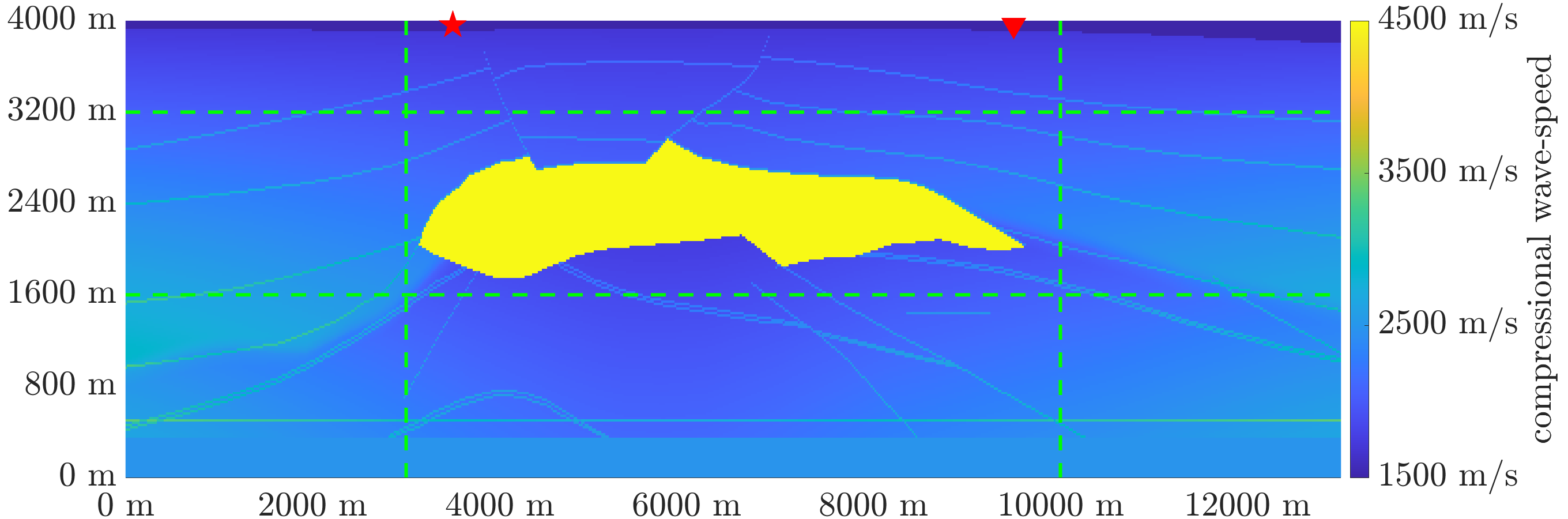}
\caption{
Illustration of the heterogeneous media (compressional wave-speed) used in Example 3.
The source and receiver locations are indicated by the star and the triangle, respectively, with a distance of 6000~m in between. 
The green dashed lines mark the interfaces that separate different subdomains. 
}
\label{Example_3_Salt_Model}
\end{figure}

In this example, we validate the proposed interface treatment with heterogeneous media.
Specifically, we consider a vertical section of the 3D SEG/EAGE salt model \textcolor{black}{\citep{aminzadeh1994seg}.} 
The compressional wave-speed is displayed in Figure \ref{Example_3_Salt_Model}, where the high velocity object in the center represents a salt body.
This 2D domain is then divided into nine subdomains, with the subdomain in the center that encapsulates the salt body being treated as the elastic region and the remaining subdomains being treated as the acoustic region.

Since the original SEG/EAGE salt model contains only the compressional wave-speed information, we supplement the density and shear wave-speed information as follows. 
Starting at the bottom of the water layer, the density follows a linear function that increases with depth from 1000~kg/m$^{3}$ to 2500~kg/m$^{3}$. 
For the subdomain in the center, the shear wave-speed value is assigned based on the compressional wave-speed value using the following range-wise formula
\begin{equation}
\label{Range_wise_formula}
c_s = 
\left\{
\begin{array}{llcll}
\\[-3.25ex]
\displaystyle 1000 &\text{ if\enskip } \!\!\!\!&c_p&\!\!\!\! \leq 2200&\!\!\!\!\!; \\[0.5ex]
\displaystyle \nicefrac{c_p}{2.15} &\text{ if\enskip } 2200 < \!\!\!\!&c_p&\!\!\!\! \leq 3500&\!\!\!\!\!; \\[0.5ex]
\displaystyle \nicefrac{c_p}{1.75} &\text{ if\enskip } 3500 < \!\!\!\!&c_p&\!\!\!\! &\!\!\!\!\!, \\%
\end{array}
\right.
\end{equation}
where $c_p$ and $c_s$ stand for compressional and shear wave-speed values, respectively, both in unit m/s.
The remaining subdomains are not assigned with shear wave-speed values since they are treated as the acoustic region.
Over the entire simulation domain, the minimal and maximal wave-speed values are 1000~m/s and 4482~m/s, respectively. 

With the media parameters described above, we are now ready to set up the numerical experiments. 
Specifically, we use the Ricker wavelet with 16~Hz central frequency and 0.125~s time delay as the source temporal profile. 
The maximal frequency in the source content is counted as 40~Hz, which, together with the minimal wave-speed value 1000~m/s, leads to a minimal wavelength at 25~m.

We assign 10 grid points per minimal wavelength for the coupled simulation, which then leads to grid spacing $\Delta x =$ 2.5~m in the discretization grids.
Since the discretization grid spacing (2.5~m) is smaller than the parametric grid spacing (20~m), we interpolate the medium parameters bilinearly from the parametric grid to the discretization grids.
The time step length is chosen as $\Delta t \approx$ 3.155e-4~s, which \textcolor{black}{corresponds} to a Courant number 0.8. 
Duration of the simulation is set to 6~s.

The \textcolor{black}{free-surface} boundary condition, i.e., $p=0$, is again considered for all exterior boundaries so that we can verify the energy-conserving property of the proposed interface treatment. 
Finally, the source and receiver locations are depicted in Figure \ref{Example_3_Salt_Model}, with a distance of 6000~m in between. 

\begin{figure}[H]
\captionsetup{width=1\textwidth, font=small,labelfont=small}
\centering\includegraphics[scale=0.2]{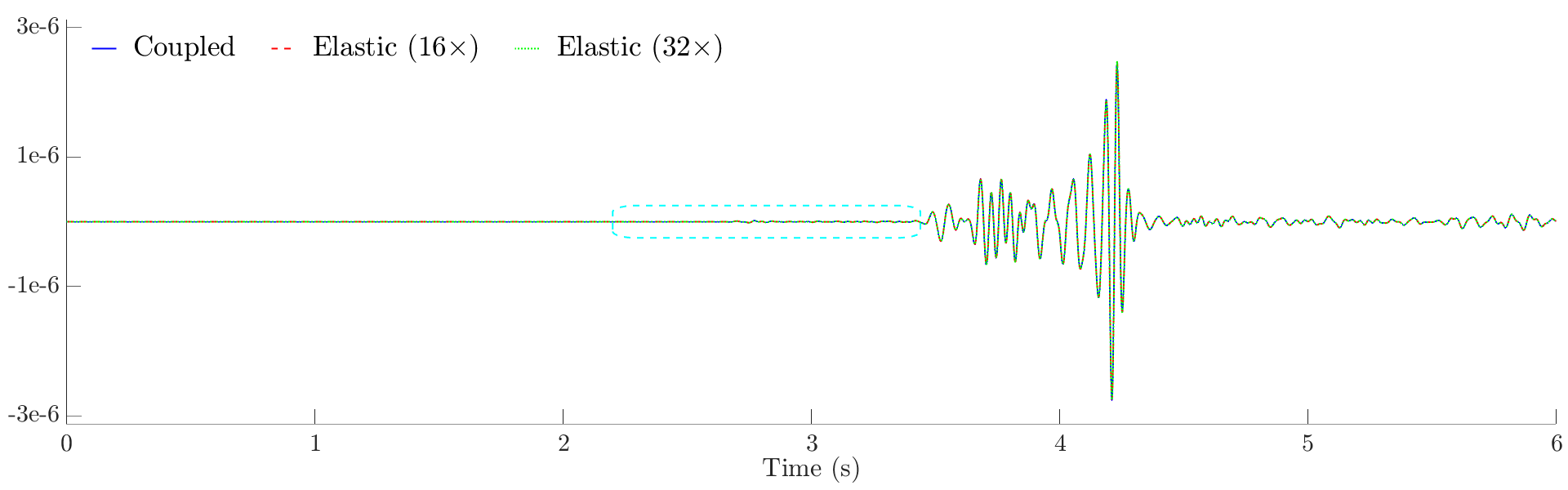}
\vspace{-.5em}
\caption{Time history of $P$ recorded at the receiver location from coupled simulation compared with full elastic simulation results.}
\label{Example_3_Seismogram}
\end{figure}

\begin{figure}[H]
\captionsetup{width=1\textwidth, font=small,labelfont=small}
\centering\includegraphics[scale=0.2]{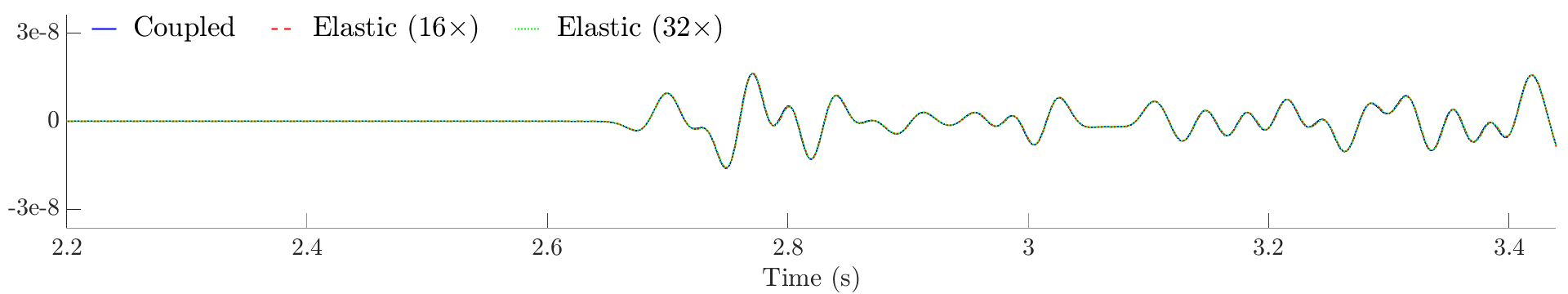}
\vspace{-.5em}
\caption{Zoom-in display of the segment encircled by the dashed lines in Figure \ref{Example_3_Seismogram}.}
\label{Example_3_Seismogram_zoom_in}
\end{figure}

As in the previous example, the recorded time history of $P$ at the receiver location from the coupled simulation is displayed in Figure \ref{Example_3_Seismogram} and compared with the full elastic simulation results. 
Zoom-in of the segment encircled by the dashed lines is displayed in Figure \ref{Example_3_Seismogram_zoom_in}, which contains waves that arrive earlier than the main wave train in Figure \ref{Example_3_Seismogram}.

\begin{figure}[H]
\captionsetup{width=1\textwidth, font=small,labelfont=small}
\centering\includegraphics[width=0.6\textwidth, scale=0.2]{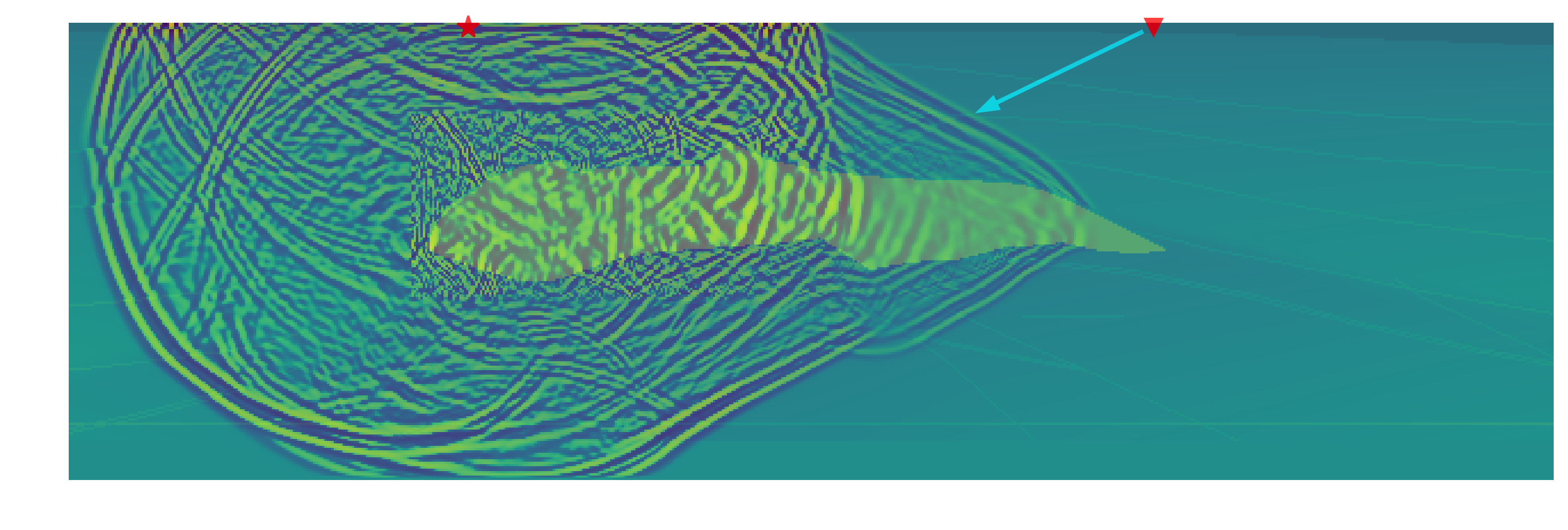}
\vspace{-.5em}
\caption{Snapshot of the wave-field (specifically, its vertical velocity component) at around 1.89~s. It can be observed that the early arrival waves, pointed at by the arrow, `radiate' from the salt body.}
\label{Example_3_Snapshot}
\end{figure}

As evidenced by the snapshot shown in Figure \ref{Example_3_Snapshot}, these early arrival waves travel through the salt body, which is the fast pathway due to the high wave-speeds associated with the salt body.
Despite their relatively small amplitude, these early arrival waves are essential for salt body characterization and sub-salt imaging.

\begin{figure}[H]
\captionsetup{width=1\textwidth, font=small,labelfont=small}
\centering\includegraphics[scale=0.2]{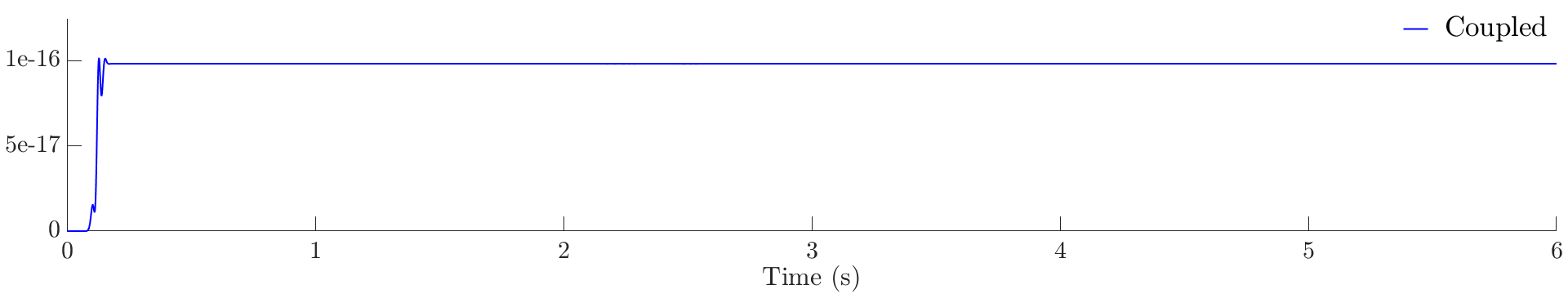}
\vspace{-.5em}
\caption{Evolution of the total discrete energy associated with the coupled simulation.}
\label{Example_3_Energy}
\end{figure}

The well matched signals in Figures \ref{Example_3_Seismogram} and \ref{Example_3_Seismogram_zoom_in} validate the accuracy of the proposed interface treatment.\footnotemark[8]
Moreover, evolution of the total discrete energy associated with the coupled simulation is displayed in Figure \ref{Example_3_Energy}, which remains constant after the source term tapers off, and hence corroborates the energy-conserving property of the proposed interface treatment. 
\footnotetext[8]{Appendix \ref{Additional_figures} contains two figures that are the equivalents of Figures \ref{Example_3_Seismogram} and \ref{Example_3_Seismogram_zoom_in}, respectively, with the receiver location placed just outside the top-right corner of the elastic region.}

\begin{figure}[H]
\captionsetup{width=1\textwidth, font=small,labelfont=small}
\centering\includegraphics[scale=0.2]{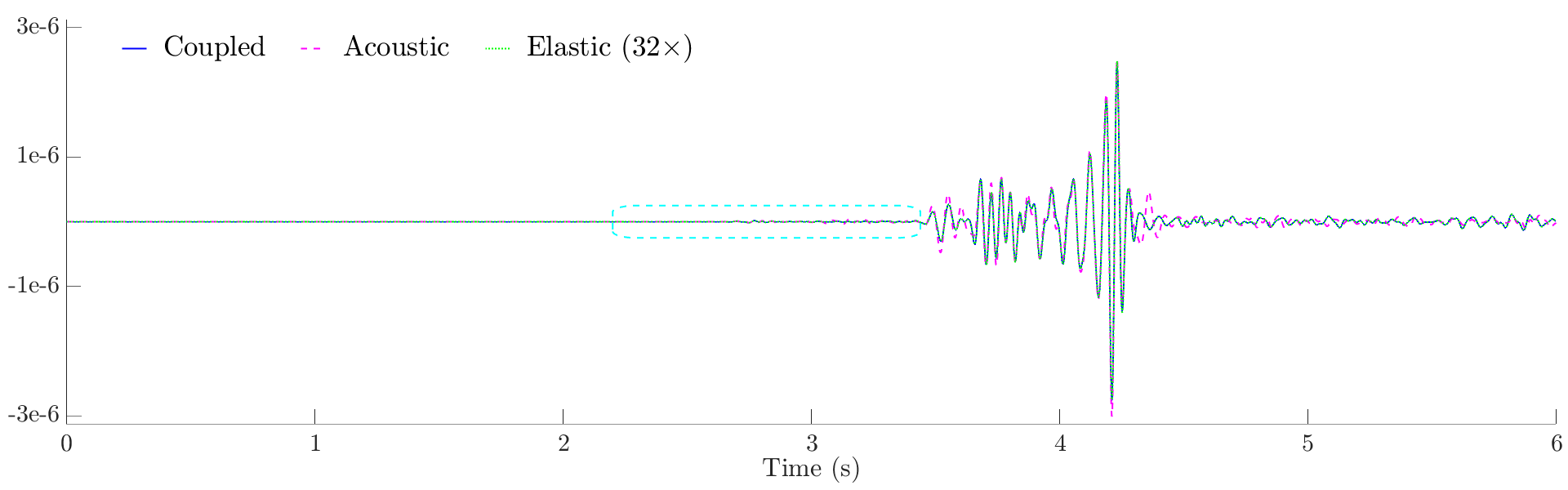}
\vspace{-.5em}
\caption{\textcolor{black}{Time history of $P$ recorded at the receiver location from coupled, full acoustic and full elastic simulation results.}}
\label{Example_3_Seismogram_with_acoustic}
\end{figure}

\begin{figure}[H]
\captionsetup{width=1\textwidth, font=small,labelfont=small}
\centering\includegraphics[scale=0.2]{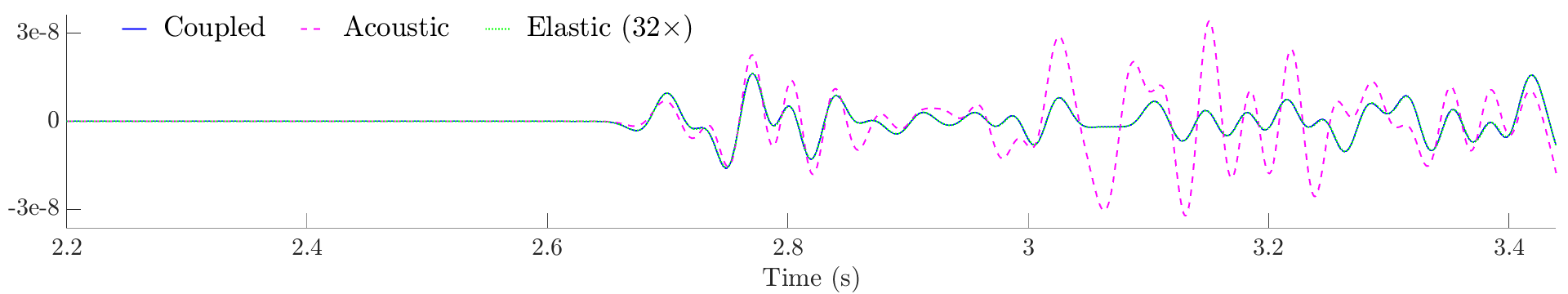}
\vspace{-.5em}
\caption{\textcolor{black}{Zoom-in display of the segment encircled by the dashed lines in Figure \ref{Example_3_Seismogram_with_acoustic}.}}
\label{Example_3_Seismogram_with_acoustic_zoom_in}
\end{figure}

\textcolor{black}{
As comparison, the seismogram simulated with the full acoustic model is displayed in Figure \ref{Example_3_Seismogram_with_acoustic}, with the zoom-in of the early arrival waves displayed in Figure \ref{Example_3_Seismogram_with_acoustic_zoom_in}.
Significant discrepancy can be observed in the early arrival waves, since the acoustic model is not able to properly capture the elastic properties of the salt body.
}

\section{Discussion} \label{Section_discussion}
The above numerical examples validate the efficacy of the presented elastic-acoustic interface treatment and exhibit the potential of the coupled elastic-acoustic wave simulations for seismic imaging applications. 
However, this study is only the first step toward a concrete strategy to properly exploit the benefits of such coupled simulations in seismic imaging. 
A potential future extension is to use different discretization grid spacings for the elastic and acoustic regions, which will lead to nonconforming grid interfaces. 
Paired interpolation operators that satisfy special relations (see, for example, equation 15 of \cite{mattsson2010stable} or equation 58 of \cite{gao2019sbp}) are required to properly address nonconforming grid interfaces.
Assisted by modern symbolic computing software, such interpolation operators can be derived without much difficulty.
Extension to coupling nonconforming elastic-acoustic interfaces can then be achieved through straightforward modifications to the penalty terms presented above using these interpolation operators, similar to what has been presented in \cite{gao2019sbp} for the acoustic case. 
\textcolor{black}{
Additionally, similar discretization technique also based on staggered grid SBP finite-difference operators has recently been extended to curvilinear grids for the acoustic wave equation \citep{o2020energy}. 
This indicates that similar extension may also be attainable for the coupled acoustic elastic wave system, which can be applied to address the irregular geometry at the ocean bottom.}

\section{Conclusion} \label{Section_conclusion}
The problem of explicitly coupling the simulations of acoustic and isotropic elastic wave systems is considered. 
Such coupled simulation has applications in seismic exploration studies involving salt bodies, 
where the region surrounding the salt body can be modeled with the elastic wave system to account for the converted waves, while the background region can still be modeled with the computationally more affordable acoustic wave system. 
Energy-conserving numerical discretization based on the techniques of summation-by-parts \textcolor{black}{finite-difference} operators and simultaneous approximation terms has been derived for the coupled wave system.
Specifically, the elastic-acoustic interface conditions are addressed by carefully designed simultaneous approximation terms that enforce these conditions weakly.
Satisfactory simulation results have been demonstrated on both piece-wise homogeneous and generally heterogeneous media.

\section{Acknowledgments} \label{Section_acknowledgments}
The authors thank the editors and referees for carefully reviewing this article and providing valuable suggestions. 
Additionally, the authors would like to thank Thierry-Laurent Tonellot and Vincent Etienne (Saudi Aramco) for helpful discussions. 
For computer time, this research used the resources of the KAUST Supercomputing Laboratory. 
This research was funded by KAUST under the Extreme Computing Research Center, KAUST grant OSR-2019-CCF-3666.4, Saudi Aramco grant RGC/3/3226-01-01, and UT Austin.

\append[Remark_implementation]{Practical implementation}
As mentioned in the section \nameref{Section_problem_description}, the forms of the continuous acoustic and elastic wave systems presented in \textcolor{black}{equations} \ref{Acoustic_wave_equation_PDT_2D_analysis} and \ref{Elastic_wave_equation_PDT_2D_analysis} are more convenient for energy analysis and formula derivation, while their equivalent forms shown in \textcolor{black}{equations} \ref{Acoustic_wave_equation_PDT_2D_implementation} and \ref{Elastic_wave_equation_PDT_2D_implementation}, respectively, are more suited for the actual implementation. 

\begin{subequations}
\label{Acoustic_wave_equation_PDT_2D_implementation}
({\normalsize Acoustic})
\begin{empheq}[left=\empheqlbrace]{alignat = 2}
\displaystyle \frac{\partial w_x}{\partial t} \enskip &= \enskip \displaystyle \frac{1}{\rho} \frac{\partial p}{\partial x}; 
\label{Acoustic_wave_equation_PDT_2D_implementation_Wx} \\
\displaystyle \frac{\partial w_y}{\partial t} \enskip &= \enskip \displaystyle \frac{1}{\rho} \frac{\partial p}{\partial y}; 
\label{Acoustic_wave_equation_PDT_2D_implementation_Wy} \\
\displaystyle \frac{\partial p}{\partial t} \enskip &= \enskip \displaystyle \rho c^2 \left(\frac{\partial w_x}{\partial x} + \frac{\partial w_y}{\partial y}\right).
\label{Acoustic_wave_equation_PDT_2D_implementation_P}
\end{empheq}
\end{subequations}

\begin{subequations}
\label{Elastic_wave_equation_PDT_2D_implementation}
({\normalsize Elastic})
\begin{empheq}[left=\empheqlbrace]{alignat = 2}
\displaystyle \frac{\partial v_x}{\partial t} \enskip &= \enskip \displaystyle \frac{1}{\rho} \left( \frac{\partial \sigma_{xx}}{\partial x} + \frac{\partial \sigma_{xy}}{\partial y} \right); \\
\displaystyle \frac{\partial v_y}{\partial t} \enskip &= \enskip \displaystyle \frac{1}{\rho} \left( \frac{\partial \sigma_{xy}}{\partial x} + \frac{\partial \sigma_{yy}}{\partial y} \right); \\
\displaystyle \frac{\partial \sigma_{xx}}{\partial t} \enskip &= \enskip \displaystyle \left(\lambda + 2\mu\right) \frac{\partial v_x}{\partial x} + \lambda \frac{\partial v_y}{\partial y}; \\
\displaystyle \frac{\partial \sigma_{xy}}{\partial t} \enskip &= \enskip \displaystyle \mu \frac{\partial v_y}{\partial x} + \mu \frac{\partial v_x}{\partial y}; \\
\displaystyle \frac{\partial \sigma_{yy}}{\partial t} \enskip &= \enskip \displaystyle \lambda \frac{\partial v_x}{\partial x} + \left(\lambda + 2\mu\right) \frac{\partial v_y}{\partial y}.
\end{empheq}
\end{subequations}

After obtaining the modified semi-discretized systems \textcolor{black}{presented in equations} \ref{Modified_Acoustic_wave_equation_Discretization_2D_analysis} and \ref{Modified_Elastic_wave_equation_Discretization_2D_analysis}, where the interface conditions have been properly accounted for, one can easily convert these systems to forms that conform with \textcolor{black}{equations} \ref{Acoustic_wave_equation_PDT_2D_implementation} and \ref{Elastic_wave_equation_PDT_2D_implementation} by viewing the appended penalty terms as modifications to the spatial derivative approximations.

For example, equation \ref{Modified_Acoustic_wave_equation_Discretization_2D_analysis_Wx} can be re-written as 
\begin{equation}
\label{Remark_equation_modified_derivative_approximation}
\mathcal A^{W_x} \boldsymbol{\rho}^{W_x} \frac{d W_x}{d t} 
\enskip = \enskip \displaystyle \ \! \mathcal A^{W_x} \widetilde{\mathcal D^P_x P},
\end{equation}
where the modified derivative approximation $\widetilde{\mathcal D^P_x P}$ is defined as 
\begin{equation}
\normalsize
\label{Modified_derivative approximation_example}
\begin{array}{rcccl}
\displaystyle \widetilde{\mathcal D^P_x P}
& \!\!=\!\! & 
\displaystyle \mathcal D^P_x P
& \!\!\! + \!\!\! &
\displaystyle \eta_A^{W_x}  \left( \mathcal A^{W_x} \right)^{-1}  \left[ \mathcal P^R_x \otimes \mathcal I^N_y \right]  \mathcal A^N_y  \left\{ \left[ \left( \mathcal E^R_x \right)^T \!\! \otimes \mathcal I^N_y\right] P - \left[ \left( \mathcal E^L_x \right)^T \!\! \otimes \mathcal I^N_y\right] \Sigma_{xx} \right\}. 
\end{array}
\end{equation}
The norm matrix $\mathcal A^{W_x}$ can now be dropped from both sides of equation \ref{Remark_equation_modified_derivative_approximation}.
Going through the same procedure for the remaining equations in system \ref{Modified_Acoustic_wave_equation_Discretization_2D_analysis}, we end up with a semi-discretized system that has a term-by-term correspondence with system \ref{Acoustic_wave_equation_PDT_2D_analysis}, which can then be converted to the form that conforms with system \ref{Acoustic_wave_equation_PDT_2D_implementation} 
by moving the coefficient matrices (e.g., $\boldsymbol{\rho}^{W_x}$ in equation \ref{Remark_equation_modified_derivative_approximation}) to the right hand side of the equations. 

Similarly, \textcolor{black}{equations}  \ref{Modified_Elastic_wave_equation_Discretization_2D_analysis} can be converted to the form that conforms with \textcolor{black}{equations} \ref{Elastic_wave_equation_PDT_2D_implementation} by following the same procedure. 
An extra complication, which stems from the tangled stress components in equations \ref{Elastic_wave_equation_PDT_2D_analysis_Sxx} - \ref{Elastic_wave_equation_PDT_2D_analysis_Syy}, is that after moving the coefficient matrices to the right hand side of the equations, the modified derivative approximation $\widetilde{\mathcal D^{V_x}_x V_x}$
will appear in the updates of both $\Sigma_{xx}$ and $\Sigma_{yy}$, as illustrated below:
\begin{subequations}
\normalsize
\label{Modified_Elastic_wave_equation_Discretization_2D_implementation}
\begin{empheq}[left=\empheqlbrace]{alignat = 2}
\label{Modified_Elastic_wave_equation_Discretization_2D_implementation_Vx}
\frac{d V_x}{d t} 
\enskip &= \enskip \displaystyle 
\left( \boldsymbol{\rho}^{V_x} \right)^{-1} 
\left( 
\widetilde{ \mathcal D^{\Sigma_{xx}}_x \Sigma_{xx} } + \mathcal D^{\Sigma_{xy}}_y \Sigma_{xy} 
\right);
\\
\label{Modified_Elastic_wave_equation_Discretization_2D_implementation_Vy}
\displaystyle \frac{d V_y}{d t} 
\enskip &= \enskip \displaystyle 
\left( \boldsymbol{\rho}^{V_y} \right)^{-1}
\left( \widetilde{ \mathcal D^{\Sigma_{xy}}_x \Sigma_{xy} } + \mathcal D^{\Sigma_{yy}}_y \Sigma_{yy} \right);
\\
\label{Modified_Elastic_wave_equation_Discretization_2D_implementation_Sxx}
\displaystyle \frac{d \Sigma_{xx}}{d t} 
\enskip &= \enskip \displaystyle 
\left( \boldsymbol{\lambda}^{\Sigma_{xx}} + 2 \boldsymbol{\mu}^{\Sigma_{xx}} \right) \widetilde{ \mathcal D^{V_x}_x V_x}
+ 
\boldsymbol{\lambda}^{\Sigma_{xx}} \mathcal D^{V_y}_y V_y;
\\
\label{Modified_Elastic_wave_equation_Discretization_2D_implementation_Sxy}
\displaystyle \frac{d \Sigma_{xy}}{d t} 
\enskip &= \enskip \displaystyle \boldsymbol{\mu}^{\Sigma_{xy}} \left( \mathcal D^{V_y}_x V_y + \mathcal D^{V_x}_y V_x \right); \\
\label{Modified_Elastic_wave_equation_Discretization_2D_implementation_Syy}
\displaystyle \frac{d \Sigma_{yy}}{d t} 
\enskip &= \enskip \displaystyle 
\boldsymbol{\lambda}^{\Sigma_{yy}} \widetilde{ \mathcal D^{V_x}_x V_x}
+ 
\left( \boldsymbol{\lambda}^{\Sigma_{yy}} + 2 \boldsymbol{\mu}^{\Sigma_{yy}} \right) \mathcal D^{V_y}_y V_y.
\end{empheq}
\end{subequations}

\append[Appendix_1D_SBP_operators]{The 1D SBP operators}
The following 1D SBP operators in equations \ref{SBP_matrices_1D_DP} - \ref{SBP_matrices_1D_AV} serve as the building blocks for the \textcolor{black}{finite-difference} discretization presented in the section \nameref{section_discretization}. 
They have already appeared in \citet{gao2019sbp} and are included here to make this work self-contained.
Interested readers may refer to \citet{gao2019sbp} for more detail.

\begin{subequations}
\label{SBP_matrices_1D}
\small
\begin{align}
& 
\label{SBP_matrices_1D_DP}
\mathcal D^N = \left[ 
\arraycolsep=3.2pt\def\arraystretch{0.75}
\begin{array}{r r r r r r r r r r r r r r r r r} 
-\nicefrac{79}{78}   &   \nicefrac{27}{26}   &   -\nicefrac{1}{26}   &   \nicefrac{1}{78}   &   \multicolumn{1}{r|}{0}      \\
 \nicefrac{2}{21}     &  -\nicefrac{9}{7}       &    \nicefrac{9}{7}     &  -\nicefrac{2}{21}   &   \multicolumn{1}{r|}{0}      \\
 \nicefrac{1}{75}     &    0			      &   -\nicefrac{27}{25}  &   \nicefrac{83}{75} &  \multicolumn{1}{r|}{ -\nicefrac{1}{25} } \\ \cline{1-5}
   &     &  \nicefrac{1}{24}  &   -\nicefrac{9}{8}   &   \nicefrac{9}{8}   &  -\nicefrac{1}{24} &  \\
   &     &    &  \nicefrac{1}{24}  &  -\nicefrac{9}{8}   &  \nicefrac{9}{8}   &  -\nicefrac{1}{24} \\
& & & & \ddots & \ddots & \ddots & \ddots & \\  
& & & & & \nicefrac{1}{24}  &  -\nicefrac{9}{8}   &  \nicefrac{9}{8}   &  -\nicefrac{1}{24} \\
& & & & & & \nicefrac{1}{24}  &  -\nicefrac{9}{8}   &  \nicefrac{9}{8}   &  -\nicefrac{1}{24} \\ \cline{8-12}
& & & & & & & \multicolumn{1}{|r}{ \nicefrac{1}{25} }  &   -\nicefrac{83}{75}   &   \nicefrac{27}{25}   &    0  &   -\nicefrac{1}{75}   \\
& & & & & & & \multicolumn{1}{|r}{0}   &    \nicefrac{2}{21}     &  -\nicefrac{9}{7}       &    \nicefrac{9}{7}     &   -\nicefrac{2}{21}   \\  
& & & & & & & \multicolumn{1}{|r}{0}   &   -\nicefrac{1}{78}     &   \nicefrac{1}{26}     &   -\nicefrac{27}{26} &    \nicefrac{79}{78} 
\end{array} 
\right];
\allowdisplaybreaks[4]
\\
&
\label{SBP_matrices_1D_DV}
\mathcal D^M = \left[ 
\arraycolsep=3.2pt\def\arraystretch{0.75}
\begin{array}{r r r r r r r r r r r r r r r r r} 
-2       &     3    &    -1    &    0    &   \multicolumn{1}{r|}{0}      \\
-1       &     1    &      0   &    0    &   \multicolumn{1}{r|}{0}     \\
\nicefrac{1}{24}  &  -\nicefrac{9}{8}    &     \nicefrac{9}{8} &   -\nicefrac{1}{24} &  \multicolumn{1}{r|}{0}  &  \\
-\nicefrac{1}{71} &   \nicefrac{6}{71}  &  -\nicefrac{83}{71}  &  \nicefrac{81}{71} &  \multicolumn{1}{r|}{ -\nicefrac{3}{71} } &  \\ \cline{1-5}
   &     &  \nicefrac{1}{24}  &   -\nicefrac{9}{8}   &   \nicefrac{9}{8}   &  -\nicefrac{1}{24} &  \\
   &     &    &  \nicefrac{1}{24}  &  -\nicefrac{9}{8}   &  \nicefrac{9}{8}   &  -\nicefrac{1}{24} \\
& & & & \ddots & \ddots & \ddots & \ddots & \\  
& & & & & \nicefrac{1}{24}  &  -\nicefrac{9}{8}   &  \nicefrac{9}{8}   &  -\nicefrac{1}{24} \\
& & & & & & \nicefrac{1}{24}  &  -\nicefrac{9}{8}   &  \nicefrac{9}{8}   &  -\nicefrac{1}{24} \\ \cline{8-12}
& & & & & & & \multicolumn{1}{|r}{ \nicefrac{3}{71} }  &  -\nicefrac{81}{71} & \nicefrac{83}{71} & -\nicefrac{6}{71}  & \nicefrac{1}{71} \\
& & & & & & & \multicolumn{1}{|r}{0} & \nicefrac{1}{24}  &  -\nicefrac{9}{8}  & \nicefrac{9}{8} & -\nicefrac{1}{24}  \\
& & & & & & & \multicolumn{1}{|r}{0} & 0  		      &  0 		           &  		      -1   &   1       \\
& & & & & & & \multicolumn{1}{|r}{0} & 0   		      &  1   			   &  	              -3   &    2   
\end{array} 
\right]; 
\allowdisplaybreaks[4]
\\
&
\label{SBP_matrices_1D_AP}
\mathcal A^N = \left[ 
\arraycolsep=6pt\def\arraystretch{0.75}
\begin{array}{r r r r r r r r r r r r r r r r r} 
\nicefrac{7}{18}   &    &    &    \multicolumn{1}{r|}{}      \\
		           &  \nicefrac{9}{8}    &     &   \multicolumn{1}{r|}{}     \\
			  &	&  1   &    \multicolumn{1}{r|}{}     \\		           
			  &    &    & \multicolumn{1}{r|}{ \nicefrac{71}{72} }  \\ \cline{1-4}
&    &    &    &  1  &    \\
&    &    &    &    &  1  &    \\
&    &    &    &    &      &  \ddots   \\
&    &    &    &    &    &    & 1 &    \\
&    &    &    &    &    &    &    & 1 &    \\ \cline{10-13}
&    &    &    &    &    &    &    &    &  \multicolumn{1}{|r}{ \nicefrac{71}{72} } \\
&    &    &    &    &    &    &    &    &  \multicolumn{1}{|r}{}   &  1  \\
&    &    &    &    &    &    &    &    &  \multicolumn{1}{|r}{}   &    &  \nicefrac{9}{8}  \\
&    &    &    &    &    &    &    &    &  \multicolumn{1}{|r}{}   &    &    &  \nicefrac{7}{18} 
\end{array} 
\right];
\allowdisplaybreaks[4]
\\
&
\label{SBP_matrices_1D_AV}
\mathcal A^M = \left[ 
\arraycolsep=6pt\def\arraystretch{0.75}
\begin{array}{r r r r r r r r r r r r r r r r r} 
\nicefrac{13}{12}   &    &     \multicolumn{1}{r|}{}      \\
		           &  \nicefrac{7}{8}    &    \multicolumn{1}{r|}{}     \\
			  &    &    \multicolumn{1}{r|}{ \nicefrac{25}{24} }  \\ \cline{1-3}
&    &    & 1 &    \\
&    &    &    & 1 &    \\
&    &    &    &    &  \ddots   \\
&    &    &    &    &    & 1 &    \\
&    &    &    &    &    &    & 1 &    \\ \cline{9-11}
&    &    &    &    &    &    &    &  \multicolumn{1}{|r}{ \nicefrac{25}{24} } \\
&    &    &    &    &    &    &    &  \multicolumn{1}{|r}{}    &  \nicefrac{7}{8}  \\
&    &    &    &    &    &    &    &  \multicolumn{1}{|r}{}    &    &  \nicefrac{13}{12} 
\end{array} 
\right].
\end{align}
\end{subequations}
We note here that the matrices presented in equation \ref{SBP_matrices_1D} correspond to the case of unit grid spacing, i.e., $\Delta x=1$. 
When applied to general cases, $\mathcal D^N$ and $\mathcal D^M$ need to be scaled by $\tfrac{1}{\Delta x}$ while $\mathcal A^N$ and $\mathcal A^M$ need to be scaled by $\Delta x$.

With the above matrices, the expression $\mathcal A^N \mathcal D^M + \left( \mathcal A^M \mathcal D^N\right)^T$ from equation \ref{1D_SBP_property} takes the following form:
\begin{equation}
\label{Q_matrix_1D}
\left[ 
\arraycolsep=5pt\def\arraystretch{0.75}
\begin{array}{r r r r r r r r r r r r r r r r} 
-\nicefrac{15}{8}  & \nicefrac{5}{4}  &  -\nicefrac{3}{8}      \\
&    &    &    &    &    \\
&    &    &    &    &    \\
&    &    &    &    &    \\ 
&    &    &    &    &    &    &    \\
&    &    &    &    &    &    &    &  \\
&    &    &    &    &    &    &    &  \\
&    &    &    &    &    &    &    & \nicefrac{3}{8}  &  -\nicefrac{5}{4}  &  \nicefrac{15}{8}
\end{array} 
\right],
\end{equation}
which can be recast as $\mathcal E^R (\mathcal P^R)^T - \mathcal E^L (\mathcal P^L)^T$ with $\mathcal E^L$, $\mathcal E^R$, $\mathcal P^L$ and $\mathcal P^R$ given by
\begin{equation}
\arraycolsep=3.2pt\def\arraystretch{0.75}
\label{Vectors_E_and_P}
\small
\mathcal E^L = 
\left[ \begin{array}{c} 1 \\ 0 \\ \vdots \\ 0 \end{array} \right]; \quad
\mathcal E^R = 
\left[ \begin{array}{c} 0 \\ \vdots \\ 0 \\ 1 \end{array} \right]; \quad
\mathcal P^L = 
\left[ \begin{array}{c} \nicefrac{15}{8} \\ - \nicefrac{5}{4} \\ \nicefrac{3}{8} \\ 0 \\ \vdots \\ 0 \end{array} \right]; \quad
\mathcal P^R = 
\left[ \begin{array}{c} 0 \\ \vdots \\ 0 \\ \nicefrac{3}{8} \\ - \nicefrac{5}{4} \\ \nicefrac{15}{8} \end{array} \right],
\end{equation}
respectively.

\append[Appendix_2D_SBP_operators]{The 2D SBP operators}
The 2D SBP operators appearing in the semi-discretized acoustic wave system \ref{Acoustic_wave_equation_Discretization_2D_analysis} are constructed from the 1D SBP operators via tensor product as follows

\vspace{-1em}
\begin{subequations}
\small
\label{Acoustic_2D_SBP_operators}
\begin{alignat}{1}
\omit
\label{Acoustic_2D_norm_matrices}
\hfill $ \mathcal A^{W_x} = \mathcal A^M_x \otimes \mathcal A^N_y, \enskip 
\mathcal A^{W_y} = \mathcal A^N_x \otimes \mathcal A^M_y, \enskip 
\mathcal A^{P} = \mathcal A^N_x \otimes \mathcal A^N_y; $ \hfill
\shortintertext{{\normalsize and}}
\omit
\label{Acoustic_2D_difference_operators}
\hfill $ \mathcal D_{x}^{W_x} \! = \mathcal D^M_x \otimes \mathcal I^N_y, \enskip 
\mathcal D_{x}^{P} \! = \mathcal D^N_x \otimes \mathcal I^N_y, \enskip
\mathcal D_{y}^{W_y} \! = \mathcal I^N_x \otimes \mathcal D^M_y, \enskip
\mathcal D_{y}^{P} \! = \mathcal I^N_x \otimes \mathcal D^N_y, $ \hfill
\end{alignat}
\end{subequations}
where $\otimes$ stands for the tensor product operation while $\mathcal I$ symbolizes a 1D identity matrix.
Constructed as such, these 2D SBP operators satisfy the following relations:

\vspace{-1.em}
\begin{subequations}
\small
\label{Acoustic_2D_SBP_relations}
\begin{align}
& \label{Acoustic_2D_SBP_relations_x} \mathcal A^P \mathcal D_x^{W_x} + \left( \mathcal A^{W_x} \mathcal D_x^P \right)^T 
\! = 
\left[ \mathcal A^N_x \mathcal D^M_x + \left( \mathcal A^M_x \mathcal D^N_x \right)^T \right] \otimes \mathcal A^N_y
 = 
\left[\mathcal E^R_x \left(\mathcal P^R_x\right)^T \! - \, \mathcal E^L_x \left(\mathcal P^L_x\right)^T \right] \otimes \mathcal A^N_y;
\\
& \label{Acoustic_2D_SBP_relations_y} \mathcal A^P \mathcal D_y^{W_y} + \left( \mathcal A^{W_y} \mathcal D_y^P \right)^T 
\! = 
\mathcal A^N_x \otimes \left[ \mathcal A^N_y \mathcal D^M_y  + \left( \mathcal A^M_y \mathcal D^N_y \right)^T \right]
 = 
\mathcal A^N_x \otimes \left[\mathcal E^R_y \left(\mathcal P^R_y\right)^T \! - \, \mathcal E^L_y \left(\mathcal P^L_y\right)^T \right], 
\end{align}
\end{subequations}
which are consequences of the 1D SBP property in equation \ref{1D_SBP_property} and the properties of the tensor product operator.

Similarly, the 2D SBP operators appearing in the semi-discretized elastic wave system \ref{Elastic_wave_equation_Discretization_2D_analysis} are constructed from the 1D SBP operators via tensor product as follows
\begin{subequations}
\small
\label{Elastic_2D_SBP_operators}
\begin{alignat}{1}
\omit
\label{Elastic_2D_norm_matrices}
\hfill $ \begin{array}{l}
\mathcal A^{V_x} = \mathcal A^M_x \otimes \mathcal A^N_y, \  
\mathcal A^{\Sigma_{xy}} = \mathcal A^M_x \otimes \mathcal A^M_y,
\\[0.75ex]
\mathcal A^{V_y} = \mathcal A^N_x \otimes \mathcal A^M_y, \ 
\mathcal A^{\Sigma_{xx}} = \mathcal A^{\Sigma_{yy}} = \mathcal A^N_x \otimes \mathcal A^N_y; 
\end{array}
$ \hfill
\shortintertext{{\normalsize and}}
\omit
\label{Elastic_2D_difference_operators}
\hfill $ \begin{array}{c}
\mathcal D_{x}^{V_x} \! = \mathcal D^M_x \otimes \mathcal I^N_y, \enskip 
\mathcal D_{x}^{V_y} \! = \mathcal D^N_x \otimes \mathcal I^M_y, \enskip 
\mathcal D_{x}^{\Sigma_{xy}} \! = \mathcal D^M_x \otimes \mathcal I^M_y, \enskip
\mathcal D_{x}^{\Sigma_{xx}} \! = \mathcal D^N_x \otimes \mathcal I^N_y,
\\[0.75ex]
\mathcal D_{y}^{V_x} \! = \mathcal I^M_x \otimes \mathcal D^N_y, \enskip 
\mathcal D_{y}^{V_y} \! = \mathcal I^N_x \otimes \mathcal D^M_y, \enskip
\mathcal D_{y}^{\Sigma_{xy}} \! = \mathcal I^M_x \otimes \mathcal D^M_y, \enskip
\mathcal D_{y}^{\Sigma_{yy}} \! = \mathcal I^N_x \otimes \mathcal D^N_y.
\end{array}
$ \hfill
\end{alignat}
\end{subequations}
Constructed as such, these 2D SBP operators satisfy the following relations:

\begin{subequations}
\small
\label{Elastic_2D_SBP_relations}
\begin{alignat}{3}
& \mathcal A^{\Sigma_{xx}} \mathcal D_x^{V_x} + \left( \mathcal A^{V_x} \mathcal D_x^{\Sigma_{xx}} \right)^T 
\!\! = \, & \left[ \mathcal A^N_x \mathcal D^M_x + \left( \mathcal A^M_x \mathcal D^N_x \right)^T \right] \otimes \mathcal A^N_y
= \, & \left[ \mathcal E_x^R \left(\mathcal P_x^R\right)^T \! - \, \mathcal E_x^L \left(\mathcal P_x^L\right)^T\!\right] \otimes \mathcal A^N_y; 
\label{Elastic_2D_SBP_relations_Sxx_Vx} \\
& \mathcal A^{V_y} \mathcal D_x^{\Sigma_{xy}} + \left( \mathcal A^{\Sigma_{xy}} \mathcal D_x^{V_y} \right)^T 
\!\! = \, & \left[ \mathcal A^N_x \mathcal D^M_x + \left( \mathcal A^M_x \mathcal D^N_x \right)^T \right] \otimes \mathcal A^M_y
= \, & \left[ \mathcal E_x^R \left(\mathcal P_x^R\right)^T \! - \, \mathcal E_x^L \left(\mathcal P_x^L\right)^T\!\right] \otimes \mathcal A^M_y;  
\label{Elastic_2D_SBP_relations_Vy_Sxy} \\
& \mathcal A^{V_x} \mathcal D_y^{\Sigma_{xy}} + \left( \mathcal A^{\Sigma_{xy}} \mathcal D_y^{V_x} \right)^T 
\!\! = \, & \mathcal A^M_x \otimes \left[ \mathcal A^N_y \mathcal D^M_y + \left( \mathcal A^M_y \mathcal D^N_y \right)^T \right]
= \, & \mathcal A^M_x \otimes \left[ \mathcal E_y^R \left(\mathcal P_y^R\right)^T \! - \, \mathcal E_y^L \left(\mathcal P_y^L\right)^T\!\right];
\label{Elastic_2D_SBP_relations_Vx_Sxy} \\
& \mathcal A^{\Sigma_{yy}} \mathcal D_y^{V_y} + \left( \mathcal A^{V_y} \mathcal D_y^{\Sigma_{yy}} \right)^T 
\!\! = \, & \mathcal A^N_x \otimes \left[ \mathcal A^N_y \mathcal D^M_y + \left( \mathcal A^M_y \mathcal D^N_y \right)^T \right]
= \, & \mathcal A^N_x \otimes \left[ \mathcal E_y^R \left(\mathcal P_y^R\right)^T \! - \, \mathcal E_y^L \left(\mathcal P_y^L\right)^T\!\right], \label{Elastic_2D_SBP_relations_Syy_Vy}
\end{alignat}
\end{subequations}
which are again consequences of the 1D SBP property in equation \ref{1D_SBP_property} and the properties of the tensor product operator.

\begin{figure}[H]
\captionsetup{width=1\textwidth, font=small,labelfont=small}
\centering\includegraphics[scale=0.1]{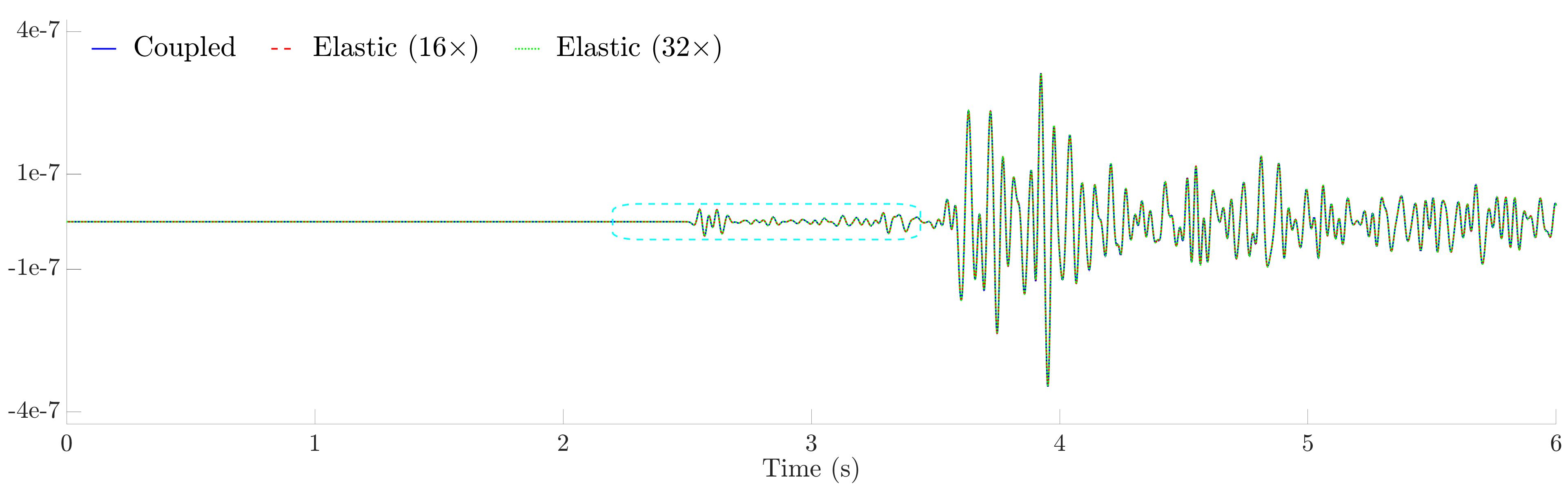}
\vspace{-.5em}
\caption{Time history of $P$ recorded at the receiver location from coupled simulation compared with full elastic simulation results.}
\label{Example_3_Seismogram_1_right_1_top}
\end{figure}

\begin{figure}[H]
\captionsetup{width=1\textwidth, font=small,labelfont=small}
\centering\includegraphics[scale=0.1]{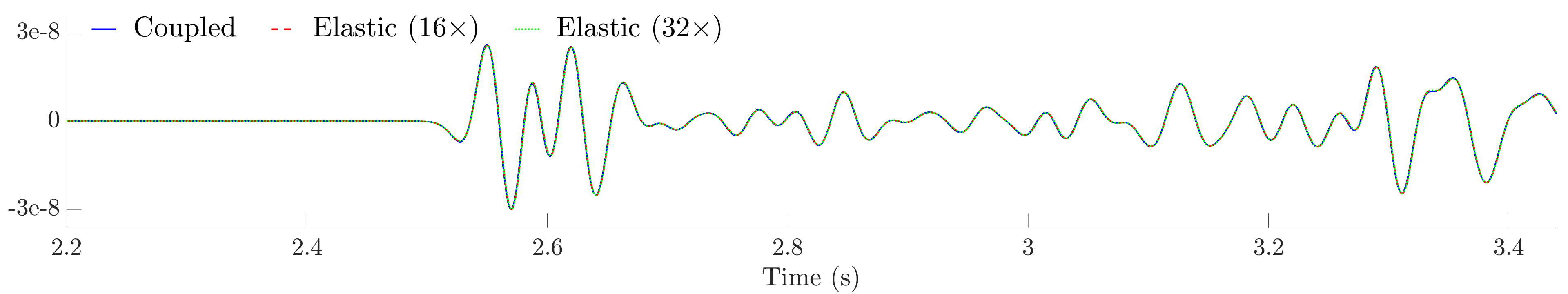}
\vspace{-.5em}
\caption{Zoom-in display of the segment encircled by the dashed lines in Figure \ref{Example_3_Seismogram_1_right_1_top}.}
\label{Example_3_Seismogram_zoom_in_1_right_1_top}
\end{figure}

\append[Additional_figures]{Additional figures}
This appendix contains two figures (\ref{Example_3_Seismogram_1_right_1_top} and \ref{Example_3_Seismogram_zoom_in_1_right_1_top}) that are the equivalents of Figures \ref{Example_3_Seismogram} and \ref{Example_3_Seismogram_zoom_in}, with the receiver location placed just outside the right-top corner of the elastic subdomain as illustrated in Figure \ref{Example_3_Salt_Model}.
Specifically, the receiver is placed at \textcolor{black}{one} parametric grid spacing (i.e., 20 m) to the top and to the right of the top-right corner. 

\bibliographystyle{seg}  
\bibliography{References}

\end{document}